\documentclass[conference]{IEEEtran}
\IEEEoverridecommandlockouts

\usepackage{cite}
\usepackage{amsmath,amssymb,amsfonts}
\usepackage{algorithmic}
\usepackage{graphicx}
\usepackage{textcomp}
\usepackage{xcolor}

\usepackage[nolist,nohyperlinks]{acronym}
\usepackage{subcaption}
\usepackage[utf8]{inputenc}
\usepackage{url}

\def\BibTeX{{\rm B\kern-.05em{\sc i\kern-.025em b}\kern-.08em
    T\kern-.1667em\lower.7ex\hbox{E}\kern-.125emX}}

\begin{document}
\title{CAVIAR: Co-simulation of 6G Communications, 3D Scenarios and AI for Digital Twins\\
    \thanks{This work was partially financed by the Innovation Center,
        Ericsson Telecomunicações S.A., Brazil;  Universal (CNPq grant
        405111/2021-5) and Project Smart 5G Core And MUltiRAn Integration
        (SAMURAI) (MC-TIC/CGI.br/FAPESP under Grant 2020/05127-2).}
}

\author{
\IEEEauthorblockN{
    João~Borges\IEEEauthorrefmark{1}, Felipe~Bastos\IEEEauthorrefmark{1}, Ilan~Correa\IEEEauthorrefmark{1}, Pedro~Batista\IEEEauthorrefmark{2}
    and~Aldebaro~Klautau\IEEEauthorrefmark{1}
}
\IEEEauthorblockA{\IEEEauthorrefmark{1}LASSE - 5G and IoT Research Group, Federal University of Pará (UFPA), Belém 66075-110 , Brazil
\\\{joao.tavares.borges, felipe.bastos\}@itec.ufpa.br, \{ilan, aldebaro\}@ufpa.br}
\IEEEauthorblockA{\IEEEauthorrefmark{2}Ericsson Research, 164 80 Stockholm, Sweden
    \\pedro.batista@ericsson.com}

}
\makeatletter
\def\ps@IEEEtitlepagestyle{%
    \def\@oddhead{\notice}
}
\def\notice{%
    {\footnotesize \textit{This work has been submitted to the IEEE for possible publication. Copyright may be transferred without notice, after which this version may no longer be accessible.}}
}

\maketitle

\begin{abstract}
    Digital twins are an important technology for advancing mobile communications,
    specially in use cases that
    require simultaneously simulating the wireless channel, 3D scenes
    and machine learning.
    Aiming at providing a solution to this demand,
    this work describes a modular co-simulation methodology called CAVIAR.
    Here, CAVIAR is upgraded to support a message passing library
    and enable the virtual counterpart of a
    digital twin system using different 6G-related simulators.
    The main contributions of this work
    are the detailed description of different CAVIAR architectures, the
    implementation of this methodology to assess a 6G use case of UAV-based
    search and rescue mission (SAR), and the generation of benchmarking data
    about the computational resource usage. For executing the SAR
    co-simulation we adopt
    five open-source solutions: the physical and link level network simulator
    Sionna, the simulator for autonomous vehicles AirSim, scikit-learn
    for training a decision tree for MIMO beam selection, Yolov8 for the
    detection of rescue targets and NATS for message passing.
    Results for the implemented SAR use case suggest that
    the methodology can run in a single machine, with the main demanded resources
    being the CPU processing and the GPU memory.
\end{abstract}


\section{Introduction}
\label{sec:intro}

The so-called \acp{DT} systems are expected to become a
prevalent technology in the future communication networks, acting as both an
enabling tool for optimizing the networks and a use case with demanding
requirements~\cite{khan2022digital}. Such systems are comprised of two elements,
called \ac{PTwin} and \ac{VTwin} \cite{wu2021digital}, where the physical twin
is the target system, which can be an object or a process under monitoring, and
the virtual twin is the simulation and the model of the physical counterpart,
reflecting its states under a given update frequency and modeling complexity
\cite{DTconsortiumDefinition}.

In the specific domain of wireless communications, \acp{DT} are increasingly
popular. For example, academic testbeds such as Colosseum \cite{bonati2021,
    villa2023colosseum, villa2023twinning} and CCI xG~\cite{cci2023}, are using a
combination of physical and virtual wireless communication environments, aiming
at developing and accurately testing new \ac{5G/6G} technologies. The industry
is also targeting \ac{5G/6G} \acp{DT}, such as Ericsson and NVIDIA working
together and using tools as Omniverse \cite{linDT2023, ericssonOmniverse}.

Among important applications of \acp{DT} in 5G / 6G, we can mention ``what-if
analysis'', network planning, generation of logs of \acp{KPI} and the support
for real-time management \cite{EricssonDigTwinWebSite, linDT2023,
    ahmadi2021networked}.

While some of these applications do not require a 3D representation of the
physical world within the simulation loop, this paper concerns a specific
category of \ac{DT} use cases in which the \ac{VTwin} is required to represent
and eventually render the world as 3D scenes. We are interested in simulating
not only the communications system but also a \emph{paired} (aligned in time)
representation of the \ac{PTwin} using 3D \ac{CGI}
\cite{shirley2009fundamentals, foley1994introduction}. Besides \ac{CGI} and
wireless communications, the simulations are also required to support
\ac{AI/ML}. The integration of these three modules require co-simulation of very
distinct software, each one specialized in its own domain: communications, 3D
\ac{CGI} and \ac{AI/ML}. This demands a proper framework and well-defined
interfaces~\cite{mecsyco,cosim20}.

We discuss here a methodology and associated software that supports this
co-simulation, called \emph{\ac{CAVIAR}}~\cite{aldebaro-ssp-2021-paper}.

\begin{figure*}[!t]
    \centering
    \begin{subfigure}[b]{0.3\textwidth}
        \centering
        \includegraphics[scale=.1365]{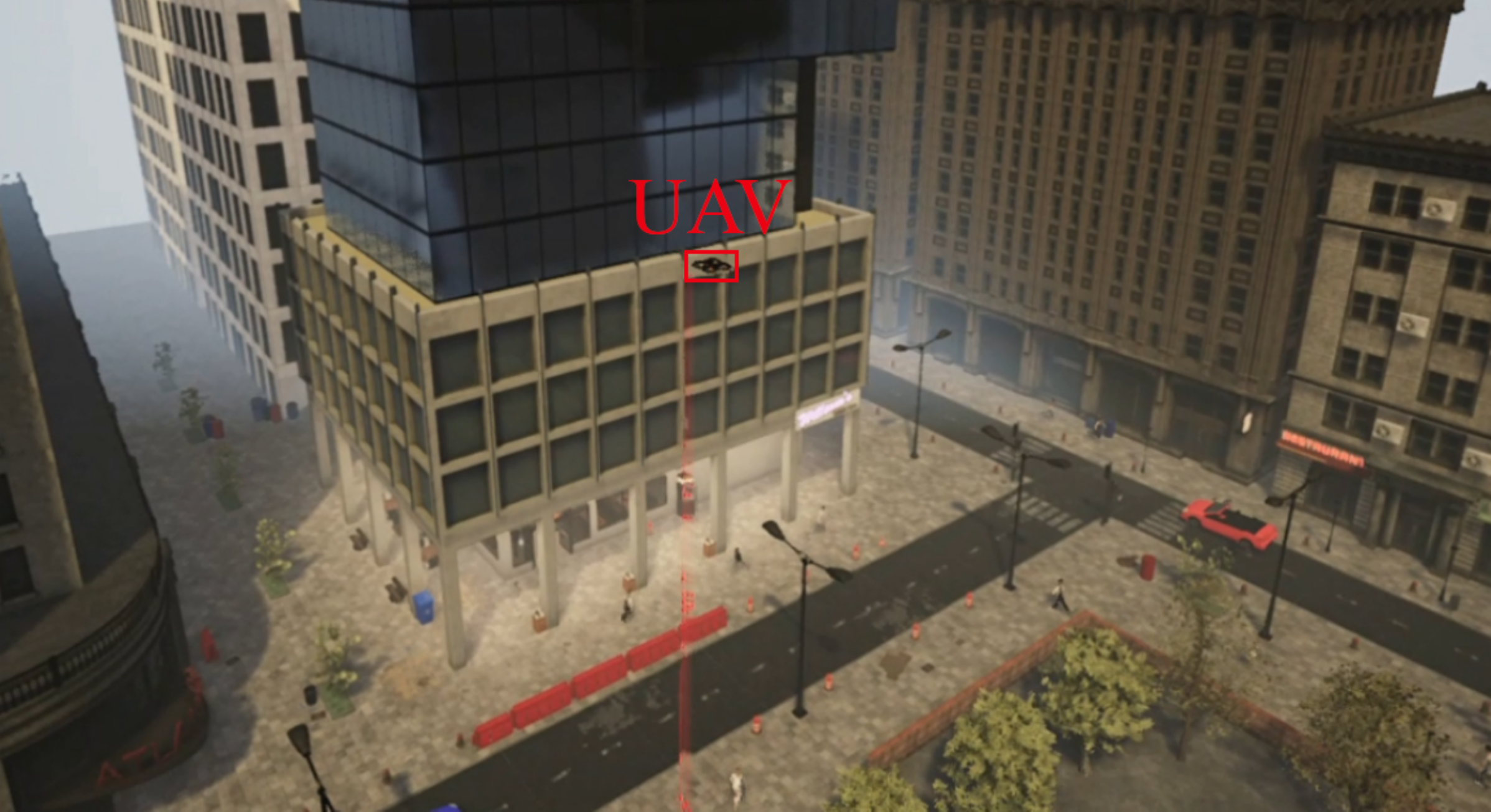}
        \caption{At time $t$.}
        \label{fig:t_0}
    \end{subfigure}
    \begin{subfigure}[b]{0.3\textwidth}
        \centering
        \includegraphics[scale=.1365]{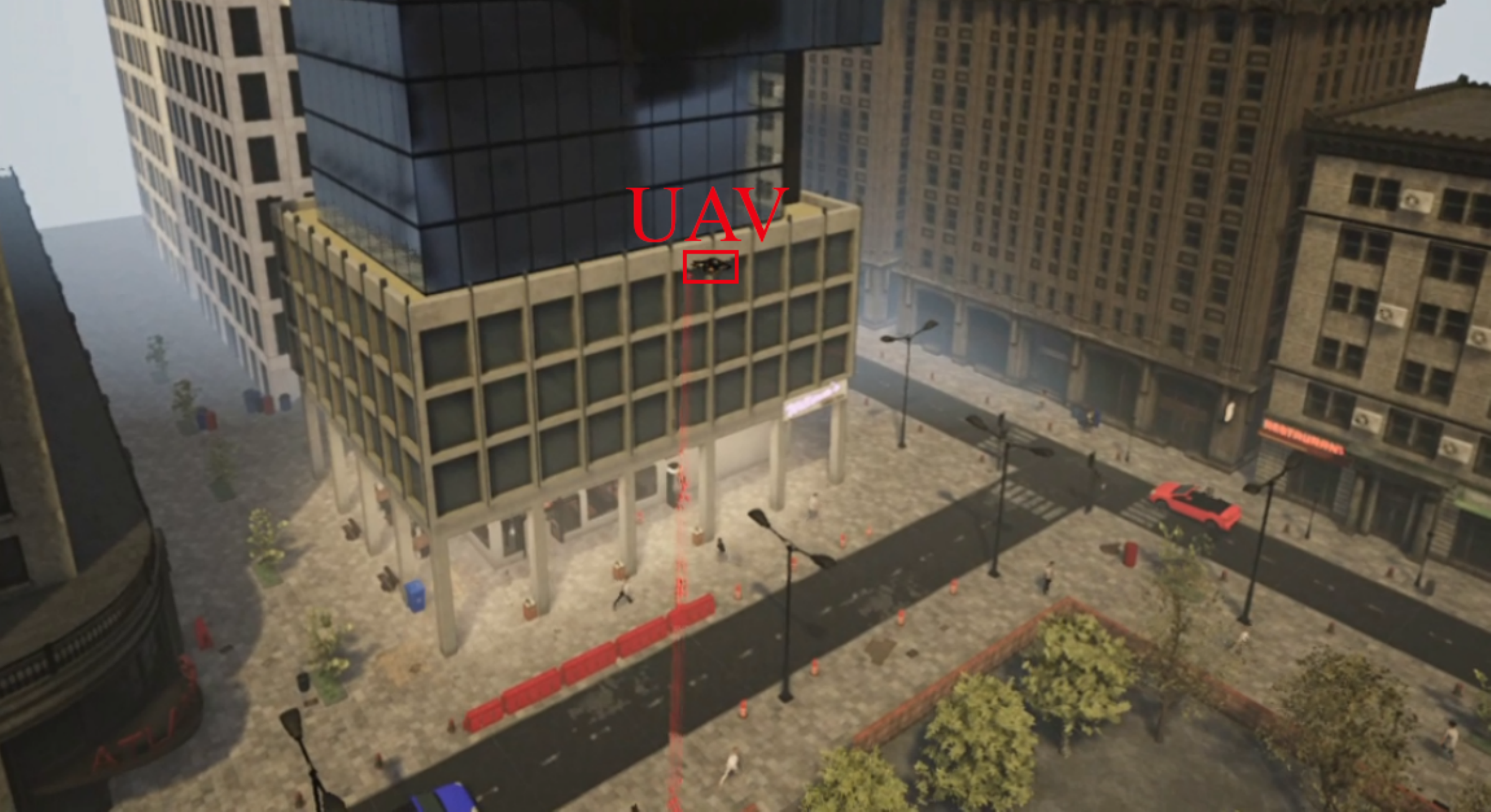}
        \caption{Time $t+ T_s$.}
        \label{fig:t_1}
    \end{subfigure}
    \begin{subfigure}[b]{0.3\textwidth}
        \centering
        \includegraphics[scale=.1365]{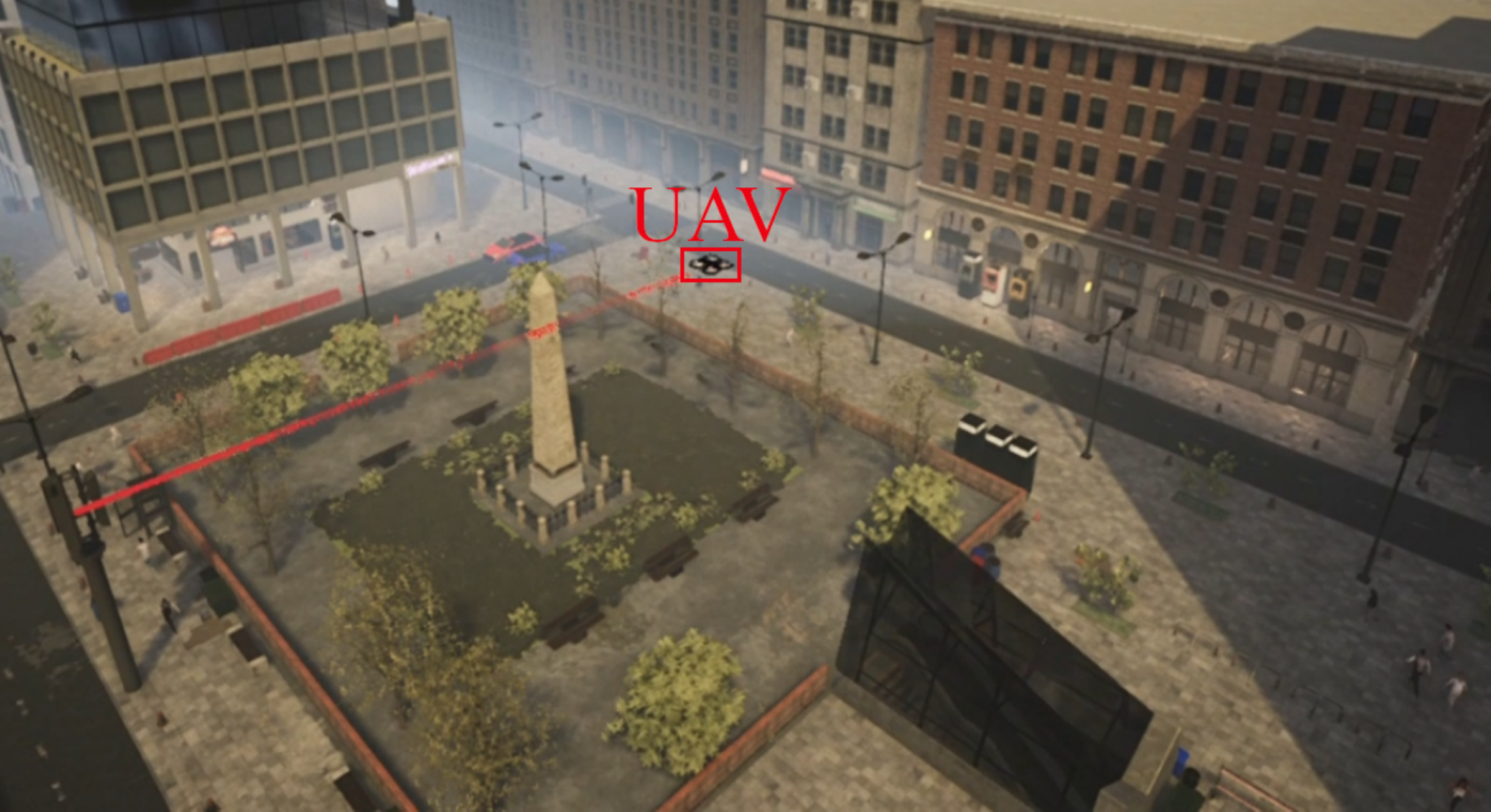}
        \caption{Time $t+ 18 T_s$.}
        \label{fig:t_muchmore}
    \end{subfigure}
    \caption{Scenes for three distinct time instants of a \ac{CAVIAR} simulation
        with an \ac{UAV} being served via beamforming. The red pointy beam is
        shown for illustrative purposes, to indicate the UAV wireless channel
        was calculated via ray tracing.}
    \label{fig:scenes_timepassing}
\end{figure*}

A \ac{DT} system with said features is an attractive tool for several \ac{5G/6G}
use cases. Two examples are \ac{ISAC}\cite{wild2021joint,liu2022integrated,
    zhang2023robust, bayraktar2023hybrid, abd2023hydra} and \ac{RIS}
\cite{demirdigital}. Both can benefit from simulations supporting wireless
communications together with detailed 3D representations when testing new
approaches. These \ac{5G/6G} uses cases can benefit from \acp{RIC} to support
\ac{AI/ML} as part of a standardized solution~\cite{abd2023hydra}. Another
motivation to a \ac{VTwin} with the \ac{CAVIAR} requirements are the many recent
papers describing how machine learning can leverage situational awareness for
reducing communication overhead (see, e.\,g., \cite{Brilhante2023}). Instead of
exclusively relying on data streamed from sensors as in \cite{Wang2018}, a
\ac{VTwin} can use \ac{CGI} via \ac{CAVIAR} to provide information from 3D
scenes (represented as simulated photos, LiDAR point clouds, etc.) to the
\ac{ML} module that aims at reducing communications overhead.

One additional example of an advanced use case where a \ac{VTwin} can greatly
benefit from a \ac{CAVIAR} simulation is \ac{SAR}: a connected \ac{UAV} is used
to search for a missing person using \ac{ML}-based object detection
\cite{lins-2021}. The \ac{UAV} trajectory is not pre-determined, but can change
according to the decisions of the \ac{ML} module. Because of that, the wireless
channels cannot be pre-computed as done, e.\,g. in~\cite{villa2023twinning}, and
the simulator needs to generate them in-the-loop of the simulation, as time
progresses and the \ac{ML} and 3D \ac{CGI} modules are invoked.

These use cases motivate our contributions. \ac{CAVIAR} is positioned to play
the role of a \ac{VTwin} to support specialized, robust, modular and flexible
simulation tools, which are integrated to work with 5G/6G use cases that
require 3D \ac{CGI}. The proposed methodology is able to deal with the highly
specialized nature of \acp{DT} systems, which are usually tailored for the
individual problems each system is initially designed to solve
\cite{liu2023systematic}.

The main contributions of this paper are:
\begin{itemize}
    \item Provide a modular co-simulation methodology, suitable for \ac{DT}
          systems, supporting the interplay between communications (including
          \ac{RT}), photorealistic 3D and \ac{AI}, while also allowing the usage
          of distinct simulators for each of the three modules;
    \item Describe its different simulation categories, highlighting that this
          work represents the first \ac{CAVIAR} version that supports
          \emph{all-in-loop} simulations, in which the \ac{ML} module can
          influence the trajectories of mobile objects and generate wireless
          channels based on such decisions;
    \item Show benchmarking results proving the feasibility of the proposed
          co-simulation methodology.
\end{itemize}

The rest of the paper is organized as follows. In
Section~\ref{sec:related_works} we highlight \ac{CAVIAR} position in the
existing literature, by pointing its main differentiating features and comparing
it with pre-existent approaches. In Section~\ref{sec:caviar_architecture}, we
describe the \ac{CAVIAR} architecture alongside its categories in terms of
simulation style. Section~\ref{sec:simu_example} describes the \ac{SAR}
\emph{blueprint} that executes all the elements that compose a \ac{CAVIAR}
simulation. In Section~\ref{sec:results}, we describe simulation results that
emphasize feasibility, showing that \ac{CAVIAR} can execute sophisticated
simulations that include \ac{ML} model inference as part of the simulation loop,
in a reasonable amount of time while not requiring specialized hardware.
Section~\ref{sec:conclusions} concludes the paper and lists future work.

\section{\ac{CAVIAR} and Related Works}
\label{sec:related_works}

\ac{CAVIAR} consists of a co-simulation methodology and associated software,
implemented as an open source Python package.\footnote[1]{The project code will
    be made available upon publication at
    https://github.com/lasseufpa/caviar.} An early version of \ac{CAVIAR} was
proposed in~\cite{aldebaro-ssp-2021-paper}. It was then further elaborated by
the authors in~\cite{oliveira2021simulation} and~\cite{borges2021reinforcement}.

As mentioned, a \ac{CAVIAR} simulation relies on three main modules: a)
\emph{3D} and its submodule \emph{mobility} supporting 3D \ac{CGI} and a
mobility simulator, b) \emph{Communications}, that plays the role of the
communications system, which may eventually include wireless and wireline links
such as a millimeter wave radio and fiber-based fronthaul, and c) the
\ac{AI}/\ac{ML} module, called \emph{\ac{AI}} for simplicity. These modules are
implemented with simulators that were independently designed, and aim at
operating in different domains. Hence, \ac{CAVIAR} adopts a Python
\emph{Orchestrator} to integrate the adopted simulators.

\begin{figure}[!htb]
    \centerline{\includegraphics[scale=.6]{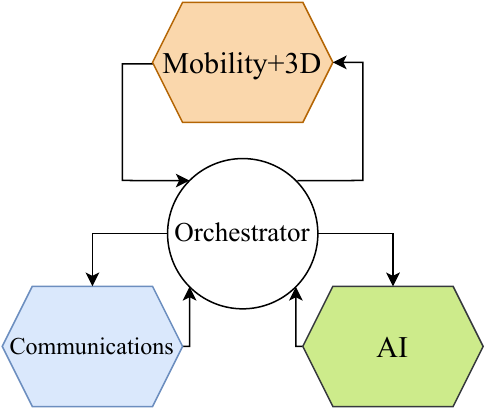}}
    \caption{Representation of the three main \ac{CAVIAR} modules and their
        interconnection via the orchestrator.}
    \label{fig:CAVIAR_modules}
\end{figure}

The Orchestrator handles communications and data flow among distinct software
and physical assets, increasing simulation realism and flexibility~\cite{mecsyco,cosim20}.
Fig.~\ref{fig:CAVIAR_modules} depicts the main \ac{CAVIAR} modules and
their integration.
The two most important Orchestrator
submodules are the ones responsible for \emph{scheduling} and \emph{message
    passing}.

Scheduling the events is an important task and
its detailed description is out of the scope of this paper.
For simplicity, here it will be assumed that the simulators
are periodically invoked sequentially, according to a virtual discrete time
determined by the sampling instant $T_s$. For example, Fig.~\ref{fig:scenes_timepassing} depicts
three \emph{scenes} at time instants $t$, $t+T_s$ and $t + 18 T_s$ seconds,
where $T_s=500$~ms. Another key component is message passing, which can leverage open source
libraries such as \ac{NATS} and \ac{0MQ}.

An important characteristic is that \ac{CAVIAR} is not bound to using a specific
software for a given module. TABLE~\ref{tab:possible_tools} exemplifies some of
the possible simulators/tools that have been used.
The modular
approach helps the maintainability of \ac{CAVIAR} tools, and facilitate their
customization to specific \acp{DT}.

\begin{table}[htb]
    \centering
    \small
    \caption{Examples of compatible simulators/tools for each \ac{CAVIAR}
        module.}
    \begin{tabular}{c|c}
        \textbf{Module} & \textbf{Example of compatible simulators/tools} \\
        \hline
        Mobility        & \textbf{Mobility:} AirSim, SUMO                 \\
        + 3D            & \textbf{3D:} Unreal Engine, Unity3D
        \\\hline
        Communications  & NVIDIA Sionna, Wireless InSite, ns-3
        \\\hline
        AI              & TensorFlow, PyTorch, scikit-learn
        \\\hline
        Message passing & NATS, 0MQ
        \label{tab:possible_tools}
    \end{tabular}
\end{table}

A certain combination of \ac{CAVIAR} simulation tools that is integrated,
tested and made freely available is called a \textit{blueprint}. To facilitate
their usage, \ac{CAVIAR} blueprints are numbered and deployed as \acp{VM}. For
instance, Section~\ref{sec:simu_example} discusses the blueprint that addresses
the previously mentioned \ac{SAR} use case.

To properly relate \ac{CAVIAR} with previous works, it is useful to first list
its main requirements:
\begin{itemize}
    \item \textbf{R1} - \emph{A platform that co-simulates communications, 3D
              \ac{CGI} (including mobility) and \ac{AI}/\ac{ML}}. The
          co-simulation must support use cases such as the one addressed by
          the \ac{SAR} blueprint, in which the wireless channels cannot be
          pre-computed. Pre-computed data sets may be eventually used, but
          \ac{CAVIAR} is designed to support \emph{all-in-loop} generation
          of communication channels, 3D scenes and \ac{AI}/\ac{ML}
          predictions or model training, further explained in Subsection
          \ref{subsec:simulation_categories}.

    \item \textbf{R2} - \emph{Support to photorealism and 3D representation of
              the physical world}. A simplistic representation of the \ac{PTwin}
          may suffice in some use cases, but \ac{CAVIAR} supports
          photorealistic 3D scenes via modern engines such as Unreal
          Engine\cite{UnrealWebSite} and Unity3D \cite{UnityWebSite}.

    \item \textbf{R3} - \emph{Support to generation of site-specific
              communication channels via ray tracing}. Stochastic and hybrid
          channel models such as 3GPP 38.901 can be eventually adopted
          \cite{Pang2022, Huang2023}. However, \ac{CAVIAR} supports
          \emph{map-based} models and ray tracing \cite{Suga2023}. The
          adopted ray tracing methodology supports mobility of both
          transmitters, receivers and also scatterers~\cite{klautau20185g}.

    \item \textbf{R4} - \emph{Reasonable computational cost}. A \ac{CAVIAR}
          blueprint must have a reasonable real-time factor $\textrm{RTF}=T_p /
              T_v$, where $T_p$ is the total elapsed \emph{wall-clock}
          (``physical'') time and $T_v$ is the total simulation (``virtual'')
          time~\cite{boehm2022real}. Besides, the simulation cannot require
          specialized hardware to reach a reasonable $T_p$ (for instance, no
          longer than one hour for simulating $T_v=60$~seconds on a
          high-performance personal computer)~\cite{manalastas2023towards}.

    \item \textbf{R5} - \emph{Modularity}. \ac{CAVIAR} aims at an easy
          integration of different simulators. Instead of changing significant
          parts of the adopted simulators code to closely integrate them, the
          solution must rely on strategies such as message passing that simplify
          updating software when new versions of the simulators are released, as
          well as providing long-term support. This also facilitates replacing
          the software for a specific module, e.\,g., adopting
          OMNeT++\cite{omnetWebSite} instead of \ac{ns-3} \cite{ns3WebSite}.
\end{itemize}

From this list, one can see that some requirements are conflicting. For
instance, sophisticated 3D models are required to reach the photorealism of R2,
but this significantly increases RTF when ray tracing is adopted, which goes
against R4.

Another key requirement of \ac{CAVIAR} simulations is to harmonize the high
resolution currently supported by \ac{CGI} software with the relatively limited
resolution of 3D models supported by ray tracing software. We currently rely on
methods such as Blender's \emph{decimate} to convert ``high-poly'' 3D assets
(that are easily rendered by Unreal Engine or Unity3D) into a ``low-poly''
version that is compatible and can be processed by ray tracing software
\cite{bastoseffects}.

An additional issue is the discrepant time scales of communications and
movements. Fig.~\ref{fig:scenes_timepassing} illustrates that from time $t$ to $t
    + T_s$ the mobile objects have hardly moved. For instance, the \ac{UAV}, red and
blue cars, and pedestrians are almost at the same position. Helping
visualization was the reason for choosing $T_s=500$~ms. However, the wireless
channel can change very fast, specially with respect to the channel phase. In
many cases, the channel \emph{coherence time} is of the order of
ms~\cite{heath_book_2018}. Decreasing the interval to $T_s \le 1$~ms leads to a
long total wall-clock time $T_p$, with very similar neighboring scenes.

\begin{table*}[!t]
    \small
    \centering
    \caption{Comparison between CAVIAR simulations and related work}
    \label{tab:related_approaches}
    \begin{tabular}{c|c|c|c|c|c|c}
        \textbf{References}
                                                                                 & \textbf{Communications}
                                                                                 & \textbf{AI}                                                           &
        \textbf{\begin{tabular}[c]{@{}c@{}}Mobility\\ +3D\end{tabular}}
                                                                                 & \textbf{\begin{tabular}[c]{@{}c@{}}Supported \\ UEs\end{tabular}}     &
        \textbf{Modularity}                                                      & \textbf{\begin{tabular}[c]{@{}c@{}}Use \\
                                                                                                   case\end{tabular}}
        \\ \hline
        \begin{tabular}[c]{@{}c@{}}Raymobtime, \\ A.Klautau et. al, 2018\\
            \cite{klautau20185g}\end{tabular}
                                                                                 & \begin{tabular}[c]{@{}c@{}}Wireless \\ InSite\end{tabular}
                                                                                 & TensorFlow                                                            &
        \begin{tabular}[c]{@{}c@{}}SUMO\\ +Blender\end{tabular}
                                                                                 & Cars                                                                  &
        Unspecified                                                              & \begin{tabular}[c]{@{}c@{}}5G beam \\
                                                                                       selection\end{tabular}
        \\ \hline
        \begin{tabular}[c]{@{}c@{}}Veneris,   \\ E.Egea-Lopez et. al, 2019;\\ A.
            Ruz-Nieto et. al, 2023 \\ \cite{egea2019vehicular}
            \cite{ruz20233d}\end{tabular} & \begin{tabular}[c]{@{}c@{}}OPAL and \\
                                            OMNeT++\end{tabular}                                & Unspecified
                                                                                 & \begin{tabular}[c]{@{}c@{}}SUMO\\ +Unity3D\end{tabular}
                                                                                 & Cars                                                                  &
        Unspecified                                                              & \begin{tabular}[c]{@{}c@{}}Cooperative \\
                                                                                       automated               \\ driving\end{tabular}
        \\ \hline
        \begin{tabular}[c]{@{}c@{}}AirSimN, \\ S. Tang et. al, 2021\\
            \cite{tang2021aerodynamic}\end{tabular}
                                                                                 & ns-3
                                                                                 & PyTorch                                                               &
        \begin{tabular}[c]{@{}c@{}}AirSim\\ +Unreal\\ Engine\end{tabular}
                                                                                 & \begin{tabular}[c]{@{}c@{}}UAVs\\ + cars\end{tabular}                 &
        Unspecified                                                              & \begin{tabular}[c]{@{}c@{}}Drone \\ surveillance,
                                                                                       \\ controller \\ design, and \\ basestation \\ placement\end{tabular}                                                                             \\
        \hline
        \begin{tabular}[c]{@{}c@{}}CAVIAR, \\ A. Oliveira et. al, 2021\\
            \cite{oliveira2021simulation}\end{tabular}
                                                                                 & \begin{tabular}[c]{@{}c@{}}Statistical \\ models\end{tabular}
                                                                                 & TensorFlow                                                            &
        \begin{tabular}[c]{@{}c@{}}AirSim\\ +Unreal\\ Engine\end{tabular}
                                                                                 & UAVs                                                                  &
        Unspecified                                                              & \begin{tabular}[c]{@{}c@{}}5G beam \\
                                                                                       selection\end{tabular}
        \\ \hline
        \begin{tabular}[c]{@{}c@{}}CAVIAR, \\ J. Borges et. al, 2021\\
            \cite{borges2021reinforcement}\end{tabular}
                                                                                 & \begin{tabular}[c]{@{}c@{}}Statistical \\ models\end{tabular}
                                                                                 & PyTorch                                                               &
        \begin{tabular}[c]{@{}c@{}}AirSim\\ +Unreal\\ Engine\end{tabular}
                                                                                 & \begin{tabular}[c]{@{}c@{}}UAVs, cars\\ + pedestrians\end{tabular}    &
        Unspecified                                                              & \begin{tabular}[c]{@{}c@{}}Resource \\
                                                                                       allocation           \\  and \\ 5G beam \\ selection\end{tabular}
        \\ \hline
        \begin{tabular}[c]{@{}c@{}}CAVIAR, \\ I. Correa et. al, 2022\\
            \cite{Correa2022}\end{tabular}
                                                                                 & \begin{tabular}[c]{@{}c@{}}Statistical \\ models\end{tabular}
                                                                                 & PyTorch                                                               &
        \begin{tabular}[c]{@{}c@{}}AirSim\\ +Unreal\\ Engine\end{tabular}
                                                                                 & \begin{tabular}[c]{@{}c@{}}UAVs, cars\\ + pedestrians\end{tabular}    &
        Unspecified                                                              & \begin{tabular}[c]{@{}c@{}}Resource \\
                                                                                       allocation           \\  and \\ 5G beam \\ selection\end{tabular}
        \\ \hline
        \begin{tabular}[c]{@{}c@{}}F. Wen et. al, 2023 \\
            \cite{wen2023vision}\end{tabular}
                                                                                 & \begin{tabular}[c]{@{}c@{}}Wireless \\ InSite\end{tabular}
                                                                                 & Unspecified                                                           &
        \begin{tabular}[c]{@{}c@{}}CARLA, \\ SUMO\\ +Unreal\\
            Engine\end{tabular}                    & Cars
                                                                                 & Unspecified                                                           & \begin{tabular}[c]{@{}c@{}}5G beam \\
                                                                                                                                                               selection\end{tabular}
        \\ \hline
        \begin{tabular}[c]{@{}c@{}}L. Cazzella et. al, 2023 \\
            \cite{cazzella2023multi}\end{tabular}
                                                                                 & \begin{tabular}[c]{@{}c@{}}Wireless \\ InSite\end{tabular}
                                                                                 & Unspecified                                                           &
        \begin{tabular}[c]{@{}c@{}}MIDGARD, \\ SUMO\\ +Unreal\\
            Engine\end{tabular}                  & Cars
                                                                                 & Unspecified                                                           & \begin{tabular}[c]{@{}c@{}}Generation of \\
                                                                                                                                                               multimodal                \\ data sets for \\ communications\end{tabular}
        \\ \hline
        \begin{tabular}[c]{@{}c@{}}J. Gong et. al, 2023 \\
            \cite{gong2023scalable}\end{tabular}
                                                                                 & \begin{tabular}[c]{@{}c@{}}Stochastic \\ and deterministic \\
                                                                                       models\end{tabular}         & Unspecified
                                                                                 & \begin{tabular}[c]{@{}c@{}}GAN-based \\ models\\ +Unreal\\
                                                                                       Engine\end{tabular}            & \begin{tabular}[c]{@{}c@{}}Cars \\ +
                                                                                                                        pedestrians\end{tabular}                                      & Unspecified &
        \begin{tabular}[c]{@{}c@{}}Digital twin \\ what-if \\
            analysis\end{tabular}                                                                                                                                                                           \\
        \hline
        \begin{tabular}[c]{@{}c@{}}U. Demir et. al, 2023 \\
            \cite{demirdigital}\end{tabular}
                                                                                 & \begin{tabular}[c]{@{}c@{}}Wireless \\ InSite\end{tabular}
                                                                                 & \begin{tabular}[c]{@{}c@{}}PyTorch +\\ TensorRT\end{tabular}          &
        \begin{tabular}[c]{@{}c@{}}FLASH \\ Dataset,\\ +Blender\end{tabular}
                                                                                 & Cars                                                                  &
        Unspecified                                                              & \begin{tabular}[c]{@{}c@{}}Reflective \\
                                                                                       intelligent            \\ surfaces (RIS)\\ for maintaining \\ QoS in
                                                                                       NLOS\end{tabular}                                                                               \\ \hline
        CAVIAR (this SAR blueprint)
                                                                                 & Sionna
                                                                                 & PyTorch                                                               &
        \begin{tabular}[c]{@{}c@{}}AirSim\\ +Unreal\\ Engine\end{tabular}
                                                                                 & \begin{tabular}[c]{@{}c@{}}UAVs, cars\\ + pedestrians\end{tabular}    &
        \checkmark                                                               & \begin{tabular}[c]{@{}c@{}}UAV-based \\ search and
                                                                                       \\ rescue\end{tabular}
    \end{tabular}
\end{table*}

Given the favorable context and motivating use cases, there are several
interesting and related papers. As mentioned, prior works have not addressed all
\ac{CAVIAR} requirements but they are relevant to positioning \ac{CAVIAR} with
respect to the literature. The most relevant papers to this work are listed in
TABLE~\ref{tab:related_approaches}.

In the context of \acp{DT}, the authors in \cite{nardini2024enabling} advocate
that \acp{DT} should have simulation services available and developed a \ac{PoC}
implementation using the discrete network simulator OMNeT++, alongside its
library Simu5G \cite{nardini2020simu5g} for three \ac{DT} applications: what-if
analysis, support to decision processes, and data set generation. In
\cite{aloqaily2022integrating}, the authors outline a framework for using
\acp{DT} in the context of the metaverse, including technologies such as
\ac{VR}, \ac{AI} and 6G, but did not present any implementation. A co-simulation
approach is proposed in \cite{palmieri2022co} for \ac{CACC} use cases,
integrating a custom mobile-edge network written in Python \cite{quadri2022edge}
with Simulink \cite{simulinkWebSite}. It is also worth
mentioning~\cite{gong2023scalable}, a two-pages description of a demonstration,
in which the authors indicate the usage of \acp{GAN} based on realistic data to
generate simulations of mobile environments that react to parameter
modifications in real-time.

Also, there are published papers describing co-simulation platforms that are
related to \ac{CAVIAR} although not specifically addressing the topic of
\acp{DT}. For example, AirSim$\textsuperscript{N}$~\cite{tang2021aerodynamic}
was developed to bring communication capabilities to
AirSim~\cite{airsim2017fsr}, an autonomous vehicle simulator by combining AirSim
with \ac{ns-3}. Another important related work is the Veneris
simulator~\cite{egea2019vehicular,ruz20233d}, which is a framework that
integrates the 3D engine Unity3D, OMNeT++, and a custom ray tracing application
for estimating wireless channels. It is used for \ac{CAD}, incorporating
realistic physics into the experiments. In \cite{wen2023vision}, the Unreal
Engine and the \ac{CARLA} simulator \cite{Dosovitskiy17} are used alongside \ac{WI}
\cite{WirelessInSiteWebSite} to investigate beam selection. \ac{WI} is also used
in~\cite{demirdigital}, for studying \ac{RIS}, with Blender
\cite{BlenderWebSite} as the main software of the 3D \ac{CGI} module.

\begin{figure*}[!t]
    \centering
    \begin{subfigure}[b]{0.25\textwidth}
        \centering
        \includegraphics[scale=.50]{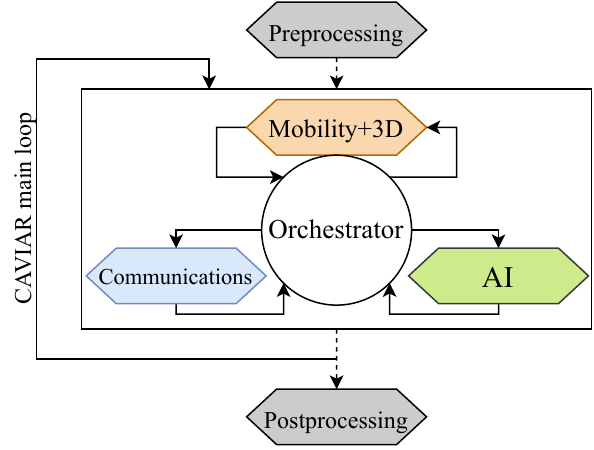}
        \caption{All-in-loop}
        \label{fig:All-in-loop}
    \end{subfigure}
    ~~~~~~~~~~~~~~~
    \begin{subfigure}[b]{0.25\textwidth}
        \centering
        \includegraphics[scale=.50]{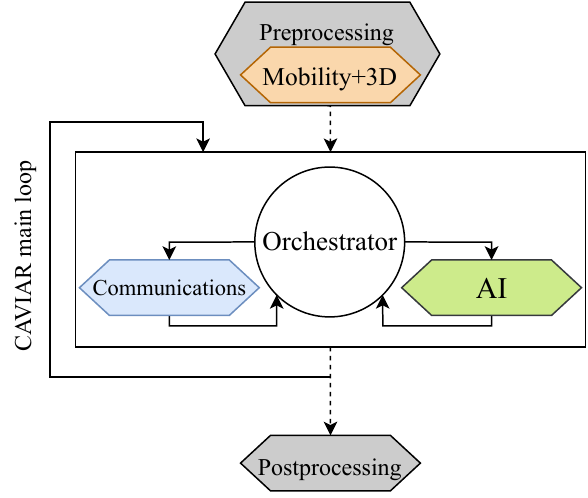}
        \caption{AI/Comm-in-loop}
        \label{fig:AI/Comm-in-loop}
    \end{subfigure}
    ~~~~~~
    \begin{subfigure}[b]{0.25\textwidth}
        \centering
        \includegraphics[scale=.50]{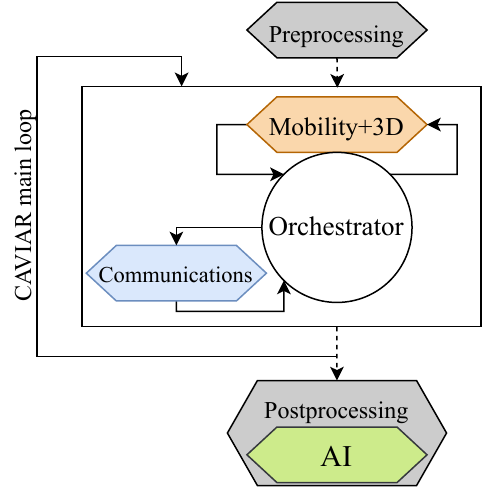}
        \caption{Mob+3D/Comm-in-loop}
        \label{fig:Mob+3D/Comm-in-loop}
    \end{subfigure}
    \caption{\ac{CAVIAR} simulation categories. (a) all-in-loop: three modules
        receive and send data in runtime; (b) AI/Comm-in-loop: \ac{AI} and
        Communications inside the loop and (c) Mob+3D/Comm-in-loop: Mobility+3D and
        Communications within the main loop.}
    \label{fig:caviar}
\end{figure*}

While AirSim$\textsuperscript{N}$~\cite{tang2021aerodynamic} and
Veneris~\cite{egea2019vehicular,ruz20233d} have similar goals, they are not as
modular as \ac{CAVIAR}. AirSim$\textsuperscript{N}$ and Veneris  present strong
dependency on specific software (AirSim and Unity3D, respectively), and lack a
clear integration with \ac{AI/ML} within the simulation loop.
TABLE~\ref{tab:related_approaches} also lists previous \ac{CAVIAR}
implementations in~\cite{oliveira2021simulation},
~\cite{borges2021reinforcement}, and ~\cite{Correa2022}, which did not report an
all-in-loop simulation and did not have a robust \textit{Orchestrator}.

Because this paper regards simulation platforms that enable users to generate
their own datasets, previous works that concern solely the release of data sets,
not the software to generate them, are out of the scope of
TABLE~\ref{tab:related_approaches}. Examples of this category are DeepMIMO
\cite{Alkhateeb2019}, DeepSense~\cite{deepsense} and radio waveform datasets
\cite{girmayIQsamples}. Besides, papers that do not integrate 3D \ac{CGI}, such
as \cite{palmieri2023co, nardini2024enabling}, and LIMoSim
\cite{sliwa2020simulating}, were also not included in
TABLE~\ref{tab:related_approaches}.

\section{\ac{CAVIAR} Architecture}
\label{sec:caviar_architecture}

Based on the literature review, CAVIAR is the first co-simulation tool for 5G/6G
\acp{DT} that is open source, modular and software agnostic, supporting: \ac{RT}
for the communication channel modeling, photorealistic 3D information, mobility
simulation, and the possibility of using \ac{AI/ML} inside the simulation loop.
This section provides details about its main components and features.

\subsection{\ac{CAVIAR} Main Loop}
\label{subsec:inloop-outloop}
\ac{CAVIAR} simulations envisions three phases: \textit{preprocessing},
\textit{main loop}, and \textit{postprocessing}, where both pre and post
processing do not need to follow strict guidelines, thus in this section we
focus on explaining the general flow of the \ac{CAVIAR} main loop.

So, at the start all the modules are initialized at $t=0$. The Mobility+3D
module instantiates the scene, composed respectively by mobile and static
elements such as vehicles and buildings. Then, the Communications module
implements its parameters, which can vary for each different simulator (e.g.
TABLE~\ref{tab:sim_params} in Section~\ref{sec:simu_example} shows an example of
the communications parameters configuration). After this, any data necessary for
the simulation is propagated among the modules at a sampling interval $T_s$ so
that, for an \emph{episode} with $N$ snapshots and duration $N T_s$, we have
$t=0, T_s, 2T_s,\cdots, (N-1) T_s$. It is assumed that a 3D scene is composed by
fixed and moving objects. Note that we adopt the term \emph{episode} as it was
used in Raymobtime~\cite{klautau20185g,raymobtime_site}, with \emph{episode}
meaning a time correlated sequence of snapshots (also called \emph{scenes}) of a
given simulation run. Positioning each mobile object at time $t$ can be done by
specialized software such as AirSim~\cite{airsim2017fsr} or
\ac{SUMO}~\cite{SUMO2018,sommer2011bidirectionally}.

\subsection{Simulation Categories}
\label{subsec:simulation_categories}
\ac{CAVIAR} simulations have different categories envisioned. In short, they
refer to adding or removing modules from inside the \ac{CAVIAR} main loop.
In~\cite{oliveira2021simulation}, these were initially called \emph{in-loop} and
\emph{out-loop}, but in this work we simplified the nomenclature as described in
the sequel.

A simulation is called \textit{all-in-loop} when all modules need to provide
data at each virtual time $t$ and can influence the next iteration of any other
module during runtime (at virtual time $t + T_s$). As suggested in
Fig.~\ref{fig:All-in-loop}, in this case all three modules are part of the
\ac{CAVIAR} main loop. This is the first paper implementing a 5G/6G
\textit{all-in-loop} co-simulation, and Sections~\ref{sec:simu_example} and
\ref{sec:results} provide details about the \textit{all-in-loop} \ac{SAR}
blueprint.

When not all modules are part of the main loop, we name the simulation category
with the module names that are part of the loop. For instance,
\textit{AI/Comm-in-loop} and \textit{Mob+3D/Comm-in-loop} simulations correspond to
Fig.~\ref{fig:AI/Comm-in-loop} and Fig.~\ref{fig:Mob+3D/Comm-in-loop}, respectively.
The AI/Comm-in-loop of Fig.~\ref{fig:AI/Comm-in-loop} exemplifies the case in
which the trajectories of mobile objects (vehicles, etc.) are pre-calculated.
Hence, the 3D scenes can be created and used to feed sensors such as cameras.

Note that the categories in Fig.~\ref{fig:caviar} do not compose an exhaustive list,
but represent some of the alternatives. For instance, a variant of
Fig.~\ref{fig:AI/Comm-in-loop} is to implement part of the Communications
processing outside the loop. One motivation is the following:
because the positions of all transceivers are known, all communication channels
can be pre-calculated (via \ac{RT}, for instance). In such cases, the \ac{AI}
can target problems as, for instance, scheduling or resource allocation, with
the pre-calculated channels being retrieved from a database, as needed.
Remaining (and faster) tasks regarding Communications can be implemented within the loop, such as
throughput and packet loss estimation. This is
similar to the situation described in~\cite{villa2023colosseum}, which uses
\emph{emulation} (with 5G protocol stacks) instead of simulation.

Fig.~\ref{fig:Mob+3D/Comm-in-loop} illustrates a \ac{CAVIAR} simulation category
that is often used for data set generation. In this case, \ac{AI} is not part of
the main loop but used at a post-processing stage. For instance, this allows
investigating \ac{MIMO} beam selection, where the served users mobility is not
influenced by the \ac{AI} module~\cite{oliveira2021simulation,
    borges2021reinforcement, Correa2022}. In spite of not having the wireless
channels generation being validated by measurements, since \cite{klautau20185g},
this strategy has extensively adopted \ac{RT} to create public channel data
sets~\cite{raymobtime_site,Alkhateeb2019, deepsense}, given the credibility of
\ac{WI} \cite{WirelessInSiteWebSite} and similar software.

\subsection{Interprocess Communication Among Modules}
\label{subsec:interprocess}

Interprocess communications are mandatory for \ac{CAVIAR} simulations, as there
is a need for signaling the execution status of the processes for an adequate
scheduling of the modules inside the main loop. However, it is out of the scope
of this paper to describe the possible distributed and heterogeneous scheduling
strategies, for which the reader is referred to \cite{boehm2022real} and
\cite{mack2022gnu}. For this work we assume that every module is executed
sequentially inside the \ac{CAVIAR} main loop.

Furthermore, besides basic scheduling, these modules need to be integrated to
share data among themselves. In this regard, \ac{CAVIAR} is based on the
strategy that, ideally, the original software should be minimally modified,
simply incorporating an interface with \ac{CAVIAR} modules. This interface
serves the purpose of sharing both, signaling messages and data. For that,
\ac{CAVIAR} adopts messaging libraries to fulfill this role, and a
\emph{publish/subscribe} pattern, which is achievable with most messaging
packages.

\begin{figure}[htb]
    \centerline{\includegraphics[scale=.6]{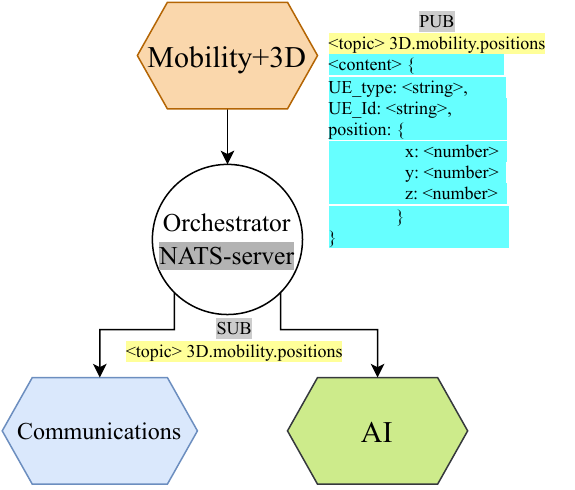}}
    \caption{Example of a publisher/subscriber dynamic within CAVIAR. In this
        case, the Mobility+3D module publishes \ac{UE} Cartesian coordinates to
        subscribers through a given topic.}
    \label{fig:CAVIAR_pubsub}
\end{figure}

In this approach, there are \emph{topics}, in which a module can interact as a
publisher or a subscriber. The role of the former is to broadcast data to the
latter. For instance, a given module can propagate the Cartesian coordinates of
the mobile elements contained within a simulation on the topic
\texttt{3D.mobility.positions}. Then, every single module interested in
receiving that information should be a subscriber to that topic, so that it can
receive and use this information in run time. This way, the modules can exchange
signals and synchronize themselves. As an illustration
Fig.~\ref{fig:CAVIAR_pubsub} shows the sharing of a \ac{UE} Cartesian
coordinates on the respective topic to any subscribing module. Also,
TABLE~\ref{tab:topics} displays the current implemented topics and the shared
information.

\begin{table}[htb]
    \centering
    \small
    \caption{Current list of topics implemented in the \ac{CAVIAR} orchestrator
        and their respective information}
    \begin{tabular}{l|l}
        \textbf{Topic}       & 3D.mobility.positions                            \\
        \hline
        \textbf{Type}        & JSON                                             \\
        \hline
        \textbf{Content}     & \begin{tabular}[c]{@{}l@{}}UE\_type
                                   \textless{}string\textgreater{}, \\ UE\_Id:
                                   \textless{}string\textgreater{}, \\ position:
                                   \textless{}JSON\textgreater{}\end{tabular}     \\
        \hline
        \textbf{Description} & \begin{tabular}[c]{@{}l@{}}UE\_type: a type used
                                   for                      \\ identification
                                   purposes.                \\ Needs to be
                                   either                   \\ ”UAV”, ”CAR” or
                                   ”PERSON”                 \\
                                   \\
                                   UE\_Id: a name used for  \\
                                   identification purposes  \\ \\
                                   position: is a JSON that \\
                                   contains numeric values  \\ for the Cartesian
                                   coordinates              \\ of the UE.
                                   Example:                 \\
                                   \{"x":0, "y":0, "z":0\}\end{tabular}  \\
        \hline \hline
        \textbf{Topic}       & communications.throughput                        \\
        \hline
        \textbf{Type}        & JSON                                             \\
        \hline
        \textbf{Content}     & \begin{tabular}[c]{@{}l@{}}UE\_type
                                   \textless{}string\textgreater{}, \\ UE\_Id:
                                   \textless{}string\textgreater{}, \\
                                   throughput
                                   \textless{}number\textgreater{}\end{tabular}
        \\
        \hline
        \textbf{Description} & \begin{tabular}[c]{@{}l@{}}The JSON contains \\
                                   identification keys           \\
                                   identically to the            \\
                                   3D.mobility.positions         \\
                                   topic. It also has a          \\
                                   key named ``throughput"       \\ that
                                   receives a number             \\
                                   and is used to propagate      \\
                                   throughput data from the      \\
                                   Communications module.\end{tabular}     \\
        \hline
        \hline
        \textbf{Topic}       & communications.state                             \\
        \hline
        \textbf{Type}        & string                                           \\
        \hline
        \textbf{Content}     & ``Ready"                                         \\
        \hline
        \textbf{Description} & \begin{tabular}[c]{@{}l@{}}Indicates the end of a
                                   \\
                                   ray tracing execution \\ for a given
                                   scene\end{tabular}
    \end{tabular}
    \label{tab:topics}
\end{table}

Now, as mentioned before, the \ac{CAVIAR} orchestration is composed of not only
signaling messages, but also of data exchange, with ideally, each \ac{CAVIAR}
module being able to support both an inward and an outward flow of data. For
this, there are two methods of data sharing among modules, which are: 1) via the
messaging library, named \textit{``message-based"} approach, or via files, named
\textit{``file-based"} approach.

The first leverages the properties of the messaging library, and the second
simply reads and writes data on easily parseable text files. Also, the inward
and outward flows can be more compatible with a message or a file-based
approach, depending on characteristics such as the data complexity (file-based)
or the need for a constant and fast-paced sharing (message-based). This is
because not all simulators have the necessary APIs or outputs message-friendly
data, such as data that can fit in a \ac{JSON}.

Finally, to observe the modularity of this approach, it is useful to consider
the case of adding a simulator to a \ac{CAVIAR} blueprint. For that, one needs
to determine: a) from which modules the added simulator will receive information
from, and b) which data it will publish to others. In case of an inward flow,
this will incur in subscribing to the topic of interest and, in case of an
outward flow, creating a topic in the format
\texttt{"module.submodule.information"}, or just \texttt{"module.information"}
and also a \ac{JSON} structure to hold the data, respectively.

So, to execute a simulation including the added software, one needs to generate
a script that imports all the necessary libraries and deals with the flow of
data, without changing much code inside the simulators themselves. This also
allows for the inclusion of scripts controlling external elements, such as the
actuators of a \ac{PTwin}, allowing the inclusion of this twin in the loop.

\begin{figure*}[t]
    \centerline{\includegraphics[scale=.5]{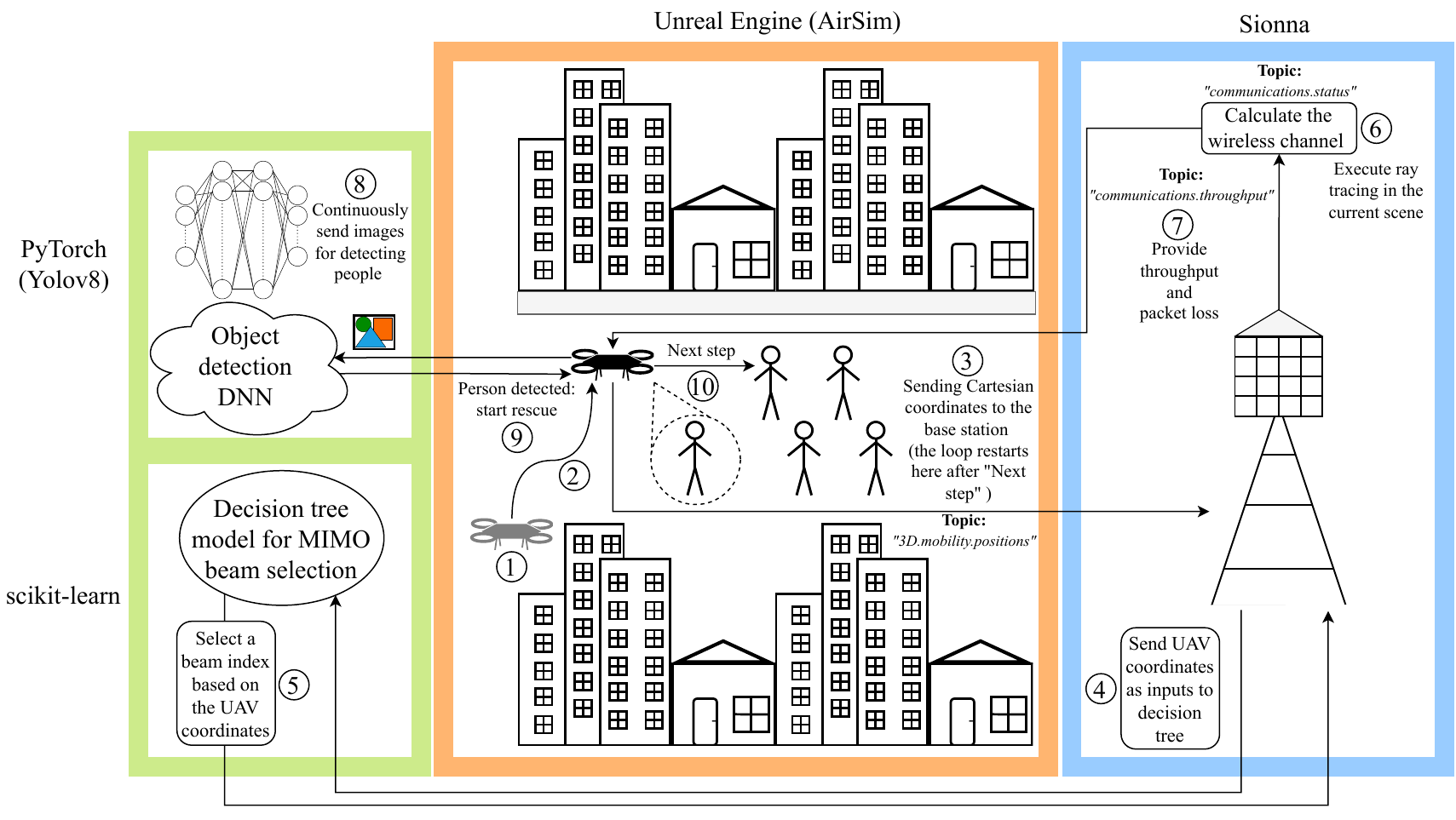}}
    \caption{The \ac{CAVIAR} blueprint for the use case presented in this work.}
    \label{fig:CAVIAR_blueprint}
\end{figure*}

\section{Search and Rescue \ac{CAVIAR} Blueprint}
\label{sec:simu_example}

This section presents the \ac{SAR} blueprint, which helps visualizing modularity
and other \ac{CAVIAR} aspects. The goal here is not to reach the best possible
\ac{ML} results, but investigate the feasibility of a co-simulation with
sophisticated and diverse simulators. The feasibility of such co-simulation is
important, given the current challenge of simulating 6G networks
\cite{manalastas2023towards}. For the \ac{SAR} blueprint, we adopted: Airsim and
Unreal Engine for the Mobility+3D module, the open-source solution
Sionna~\cite{sionna} to fulfill the role of \emph{Communications} module,
Pytorch, more specifially an \ac{YOLO} based around PyTorch (\ac{YOLO}v8) and
scikit-learn for the \ac{AI} \cite{scikit-learn}.
Fig.~\ref{fig:CAVIAR_blueprint} presents the overall system. All the used
modules are connected using \ac{NATS} as the Orchestrator's message passing
library, allowing them to work simultaneously. In the following subsections,
each adopted software is briefly detailed.

\subsubsection{Communications module}
Sionna is a physical and link layer simulator for 5G/6G systems. It allows the
quick prototyping of a communication network considering the propagation
environment, antenna patterns, carrier frequency, etc. It also has native
support for integration with \ac{ML} methods, as it is built upon the TensorFlow
library and the Python programming language.

\subsubsection{Mobility+3D module}

AirSim~\cite{airsim2017fsr} is a simulator for \acp{UAV} and cars, that
implements realistic vehicle models inside the Unreal Engine. It provides access
to all controls and sensors embedded in the vehicles, including RGB cameras,
LiDARs, barometers, etc. For this experiment, AirSim was used to control
\acp{UAV} for real-time object detection.

\subsubsection{AI module}
\ac{YOLO}v8, which is based on PyTorch, is used as an object detection \ac{DNN},
and scikit-learn is used to train a decision tree responsible for \ac{MIMO} beam
selection procedure at the \ac{BS}. These two \ac{AI} models operate on the
\ac{SAR} scenario proposed in Section~\ref{sec:simu_example}.

\subsubsection{Orchestrator}
A Python \emph{runtime} code based on \ac{NATS}\cite{nats}. \ac{NATS} is a
messaging library used for interprocess communications. With it, one can send
N-to-N messages using a ``pub/sub'' pattern as mentioned in
Subsection~\ref{subsec:interprocess}. \ac{NATS} enables the distribution of
simulations over several machines, however exploring this feature is beyond the
scope of this paper. In the \ac{SAR} blueprint, \ac{NATS} is used to exchange
data about the \ac{UE} position and the achieved throughput. In this blueprint,
the Orchestrator forces the Mobility+3D module to wait for the execution of the
\ac{RT} before moving to the other step.

\begin{figure}[htb]
    \centerline{\includegraphics[scale=.48]{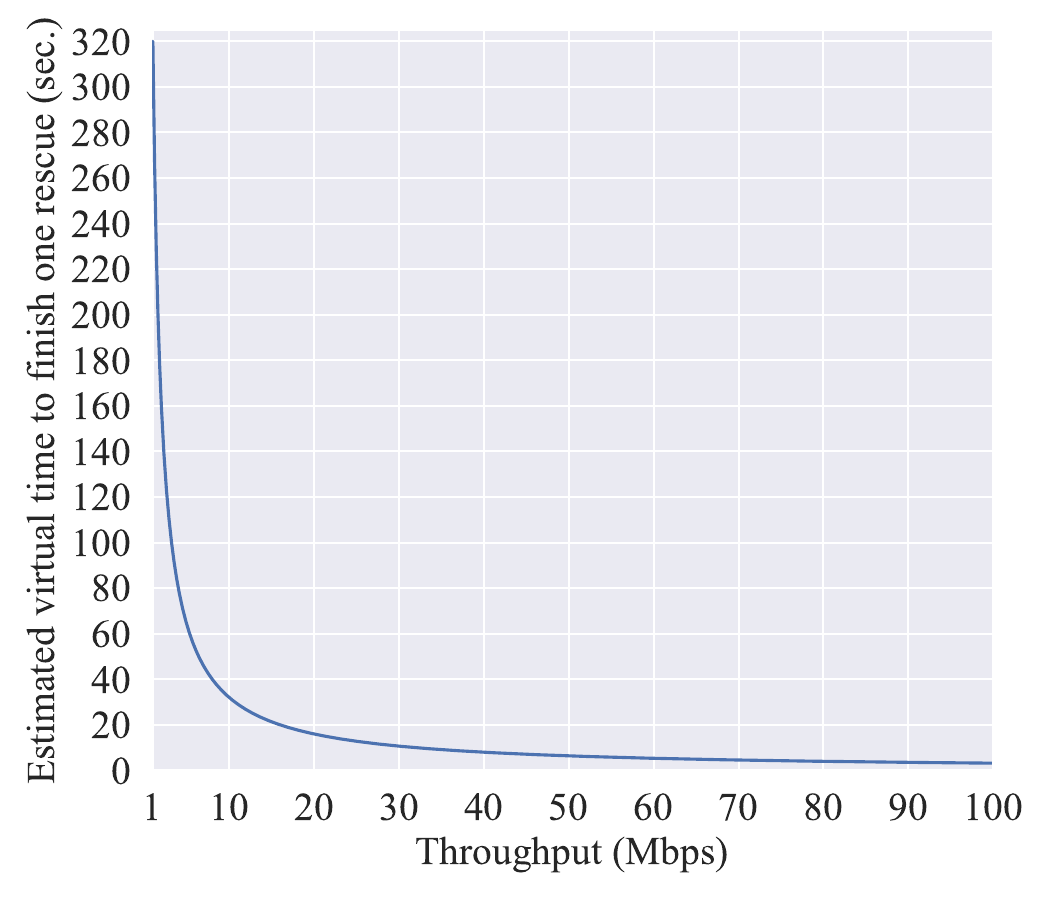}}
    \caption{Estimated virtual time to finish the rescue of one person with a given throughput value.}
    \label{fig:rescue_times}
\end{figure}

In the proposed scenario, shown in Fig.~\ref{fig:CAVIAR_blueprint}, a \ac{UAV}
carries a camera to capture video that is processed to detect victims using
computer vision based on a \ac{DNN} running on the cloud, similar to the case
discussed in \cite{lins-2021}. The goal is to execute a \ac{SAR} operation where
\ac{MIMO} beam selection is used to improve the connectivity of a \ac{UAV}
executing high quality image transmission for object detection. To simplify the
simulation, instead of incorporating \emph{path planning}, the \acp{UAV} follow
trajectories imposed by pre-determined \emph{way points}. Along these
trajectories, five ($5$) human-shaped 3D models are placed in different
positions. During the flight, the \ac{UAV} will continually send captured images
to the cloud, where the object detection \ac{ML} model will try to infer the
existence of people in the image. Once a person is detected, the drone will stop
and wait for a number $S$ of seconds, representing the time to transmit $B$
bytes of data to the mission command center. In spite of not having the
trajectory changing according to the \ac{AI} module, this behavior was
implemented as an \emph{all-in-loop} simulation (see Fig.~\ref{fig:All-in-loop})
because $S$ depends on \ac{AI}.

The mission duration will be longer or shorter depending on the throughput
achieved by the \ac{UAV}. Also, the image will suffer degradation with varying
degrees of packet loss for lower throughputs. The dynamic takes place on the
\ac{BS}, where an \ac{AI/ML} model (decision tree) is used for \ac{MIMO} beam
selection. Best beam pair choices lead to higher throughputs and shorter rescue
times, as depicted in Fig.~\ref{fig:rescue_times} for $40$ Megabytes ($B =
    4\times10^7$), corresponding to $10$ images of $4$ MB.

Regarding the image transmission over the channel, we have developed software to
transmit the bitstreams through ns-3, but this made the simulation take too
long. So, we chose to evaluate this in a followup work. As an alternative, we
implemented the following simpler mechanism to relate the loss of performance in
beam selection with a loss of quality in received image. The image degradation
due to increased packet loss provoked by bad performance in beam selection
simply followed TABLE~\ref{tab:sar_img_deg}. The image is affected by three
possible values of packet loss depending on the throughput during the rescue.
The \ac{PSNR} \cite{hore2010image} between the original and received images is
used to indicate the degradation through wireless transmission.

\begin{table}[htb]
    \centering
    \small
    \caption{Image degradation given the throughput achieved by \ac{ML}-based
        beam selection}
    \begin{tabular}{c|c|c}
        \begin{tabular}[c]{@{}c@{}}Throughput \\ range \\
            (Mbps)\end{tabular}              &
        \begin{tabular}[c]{@{}c@{}}Packet \\ loss \\ (\%)\end{tabular} &
        \begin{tabular}[c]{@{}c@{}}Approx. \\ resulting\\ PSNR (dB)\end{tabular}
        \\
        \hline
        $90 > T > 60$                                                  & $1$  &
        $\sim$$26.36$
        \\
        $60 \ge T > 30$                                                & $25$ &
            $\sim$$12.39$
        \\
        $30 \ge T \ge 0$                                               & $50$ &
        $\sim$$9.37$
    \end{tabular}
    \label{tab:sar_img_deg}
\end{table}

Two baseline approaches were used to facilitate observing the performance of the
\ac{ML}-based beam selection: \emph{random} and \emph{oracle}. The first chooses
a beam pair based on a discrete uniform distribution, with values ranging from
$[0, 3]$ for the \ac{Rx} and $[0, 63]$ for the \ac{Tx}. The oracle approach
chooses the best beam pair, which is available from the data set given that it
supports supervised learning. As depicted in Fig.~\ref{fig:CAVIAR_blueprint},
in this all-in-loop \ac{SAR} \ac{CAVIAR}
blueprint, \ac{NATS} was used to share information between the different modules
as follows: when an \ac{UAV} moves according to the mobility simulator, Sionna
updates the position of this user in the communications simulation. More
specifically, it propagates Cartesian coordinates (X, Y and Z) representing the
position of the receiver \ac{UAV} in the topic \texttt{3D.mobility.positions}.
It also propagates the current achieved bitrate in
\texttt{communications.throughput}, and finally, whether or not the ray tracing
execution finished, using the topic \texttt{communications.state}.

\begin{figure}[htb]
    \centerline{\includegraphics[width=.45\textwidth]{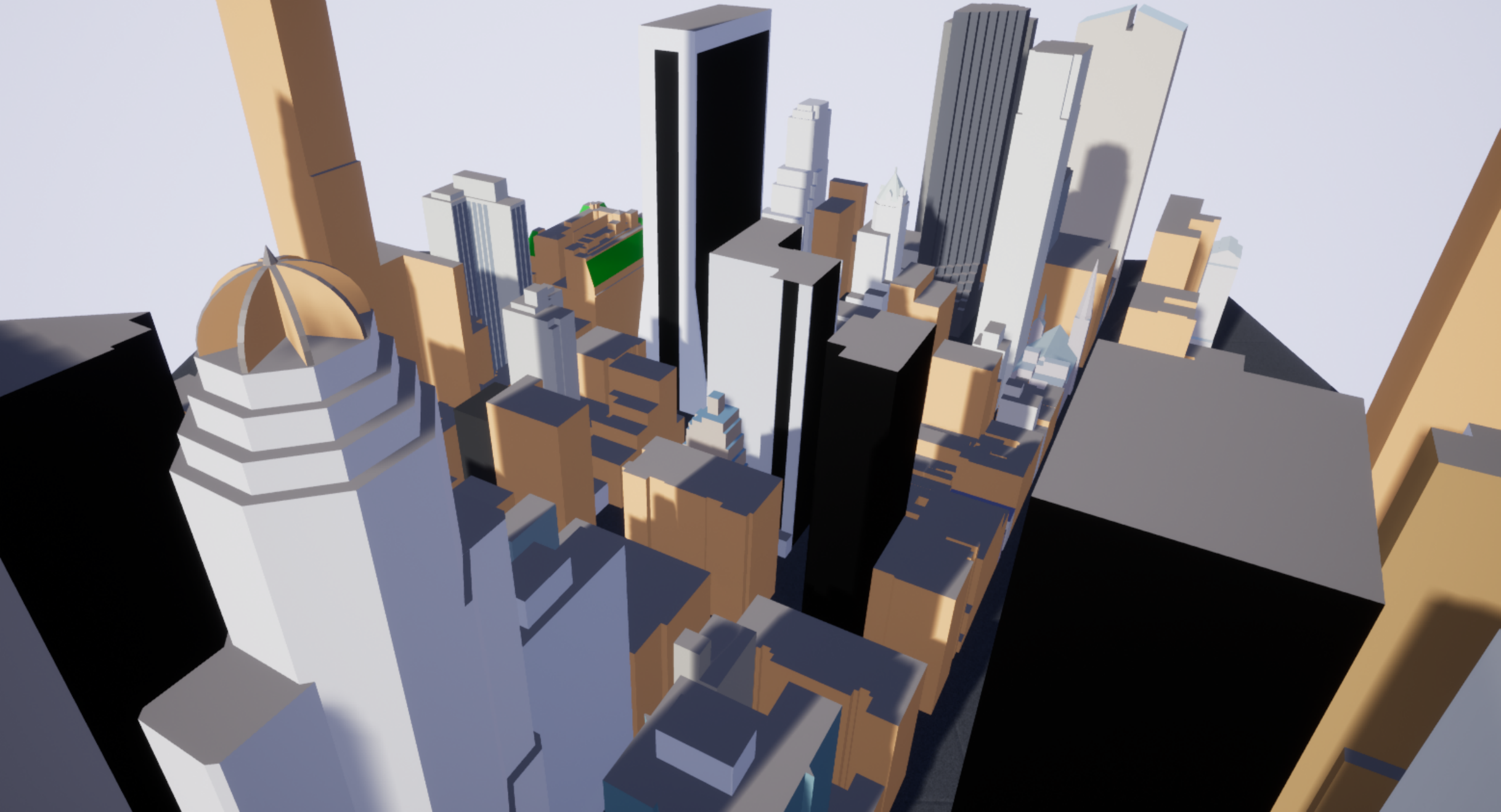}}
    \caption{Panoramic view of the Central Park scenario. The models and textures were obtained from OpenStreetMap and rendered in Unreal Engine.}
    \label{fig:central_park_unreal_panoramicview}
\end{figure}

\begin{figure}[htb]
    \centerline{\includegraphics[scale=.3]{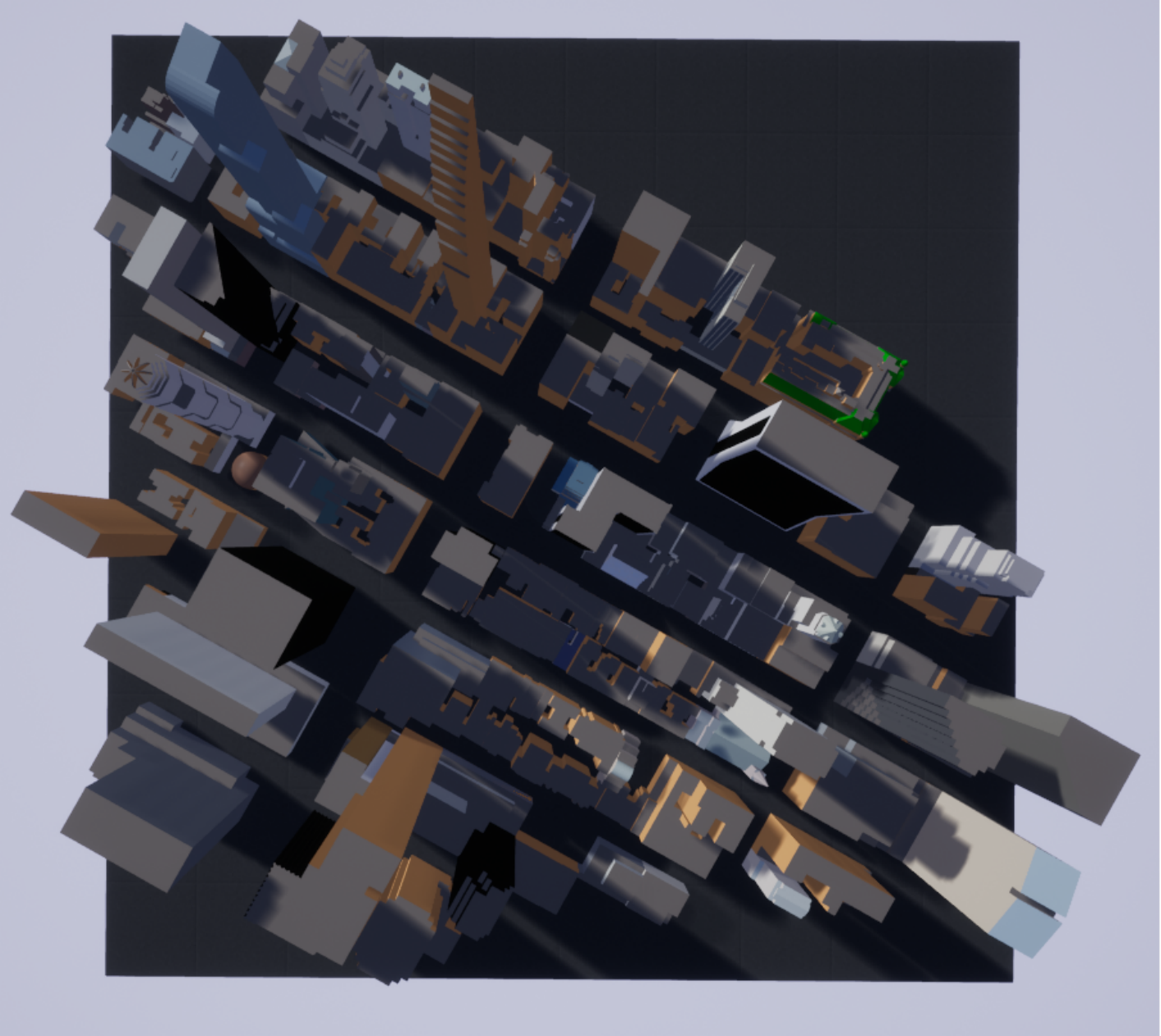}}
    \caption{Top view of the Central Park scenario in Unreal Engine.}
    \label{fig:central_park_unreal_topview}
\end{figure}

\begin{figure}[htb]
    \centerline{\includegraphics[scale=.45]{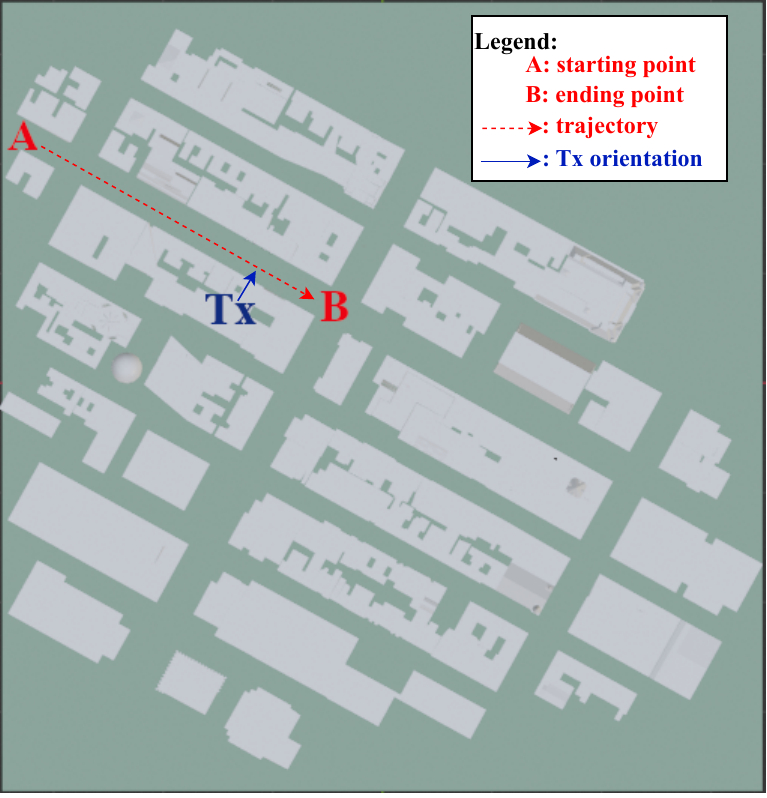}}
    \caption{Central Park scenario with the UAV waypoints for the trajectory
        from A to B and Tx positioning. The version used in Sionna does not use
        textures and underwent a slight simplification using the \textit{decimation}
        modifier in Blender.}
    \label{fig:scenario}
\end{figure}

This \ac{SAR} blueprint used a rectangular perimeter with approximately $719.2$
meters of length and $693.4$ meters of width, near the Central Park, in New
York. The model was imported from \textit{\ac{OSM}}\cite{OpenstreetmapWebSite}
and made into two versions: the original textured one, shown in
Fig.~\ref{fig:central_park_unreal_panoramicview} and
Fig.~\ref{fig:central_park_unreal_topview}, obtained directly from \ac{OSM}, for
the 3D module, and one for the Communications module, which was obtained from
the original by using Blender \cite{BlenderWebSite} with a \textit{decimation}
modifier at $0.75$ ratio, for faster ray tracing. The simplified 3D model ended
up containing a total of $9803$ faces and is exemplified in
Fig.~\ref{fig:scenario}.

The drone followed a trajectory from the point \textbf{A} to \textbf{B}, with
roughly $331$ meters, with a speed of $5$ m/s and $5$ random waypoints spread
within it to generate the decision tree train/test sets and another one called
``SAR", excluding \textbf{A} but including \textbf{B}. The \ac{Tx} was
positioned on the top of a building, at $120$ meters high, facing the street
where the drone was moving and tilted downwards in $45$ degrees. The scenario
adopted the communications parameters described in TABLE~\ref{tab:sim_params}.

\begin{table}[htb]
    \centering
    \small
    \caption{Communications parameters for the given \ac{SAR} scenario}
    \begin{tabular}{ccl}
        \hline
        \multicolumn{3}{c}{\textbf{Antenna parameters (Tx and Rx)}}   \\
        \hline
        \textbf{Type}                                               &
        \multicolumn{2}{c}{Planar array (UPA)}                        \\
        \textbf{Transmitter}                                        &
        \multicolumn{2}{c}{8x8}                                       \\
        \textbf{Receiver}                                           &
        \multicolumn{2}{c}{2x2}                                       \\
        \textbf{Vertical spacing}                                   &
        \multicolumn{2}{c}{0.5}                                       \\
        \textbf{Vertical spacing}                                   &
        \multicolumn{2}{c}{0.5}                                       \\
        \textbf{Antenna pattern}                                    &
        \multicolumn{2}{c}{TR38901}                                   \\
        \textbf{Polarization}                                       &
        \multicolumn{2}{c}{Vertical}                                  \\
        \hline
        \multicolumn{3}{c}{\textbf{Sionna scene parameters}}          \\
        \hline
        \textbf{Carrier frequency}                                  &
        \multicolumn{2}{c}{40 GHz}                                    \\
        \textbf{Radio material (building and ground)}               &
        \multicolumn{2}{c}{ITU concrete}                              \\
        \textbf{Radio material (drone)}                             &
        \multicolumn{2}{c}{ITU metal}                                 \\
        \textbf{Synthetic array}                                    &
        \multicolumn{2}{c}{True}                                      \\
        \hline
        \multicolumn{3}{c}{\textbf{Ray tracing parameters}}           \\
        \hline
        \textbf{Maximum ray bounces}                                &
        \multicolumn{2}{c}{5}                                         \\
        \textbf{Diffraction and scattering}                         &
        \multicolumn{2}{c}{Enabled}                                   \\
        \textbf{Ray tracing method}                                 &
        \multicolumn{2}{c}{Fibonacci}                                 \\
        \multicolumn{1}{l}{\textbf{Rays shot in random directions}} &
        \multicolumn{2}{c}{1.000.000}                                 \\
        \label{tab:sim_params}
    \end{tabular}
\end{table}

\section{Simulation Results}
\label{sec:results}
The results are organized in three distinct categories: 1) first the beam
selection performance of the trained \ac{AI} model, 2) the mission duration
involving the flight through the trajectories and identification of people in
need of rescue, as described in Section~\ref{sec:simu_example}, and 3) a set of
benchmarking procedures for assessing the feasibility of CAVIAR simulations with
respect to computational cost.

\subsection{Decision Tree Training and Testing}
\label{subsec:dec_tree}

The beam selection task was posed as a classification problem, with the \ac{UAV}
Cartesian coordinates as input features. We assume \ac{DFT} codebooks, with 64
codewords at the transmitter and 4 at the receiver~\cite{klautau20185g,
    heath_book_2018}. The \ac{AI} model role is to choose between an index from the
range $[0, 255]$ representing one of the possible beam pairs.

The decision tree was trained offline. For this training, a data set was
generated as follows. The Orchestrator started the flight mission and let it run
with a sampling interval of $T_s = 500$ milliseconds in virtual time. We used a
script that randomly chose way points within the trajectory show in
Fig.~\ref{fig:scenario}. At each time $t$, it stopped the simulation and sent
the \ac{UAV} coordinates to Sionna, which executes \ac{RT}. Afterwards, the
Orchestrator uses the \ac{RT} output to generate \texttt{.npz} files, containing
the \ac{UAV} position and the corresponding best beam pair index for time $t$.
This best beam pair index is obtained by using a \emph{full sweep} and returning
the two indexes that yielded the highest \emph{combined channel magnitude} among
all pairs~\cite{klautau20185g}. We used the decision tree to fit a data set
composed of $1274$ scenes from the mission.

We picked only the \ac{NLOS} channels, meaning the channels composed solely by
\acp{MPC} that were subject to reflection and diffraction/scattering on its way
from the \ac{Tx} to the \ac{Rx}. We discarded the scenes which have channels
with a \ac{LOS} \ac{MPC} because they pose a much easier task when it comes to
beam selection~\cite{klautau20185g}.

\begin{figure}[htb]
    \centerline{\includegraphics[scale=.5, trim=0 0 0 0,
            clip]{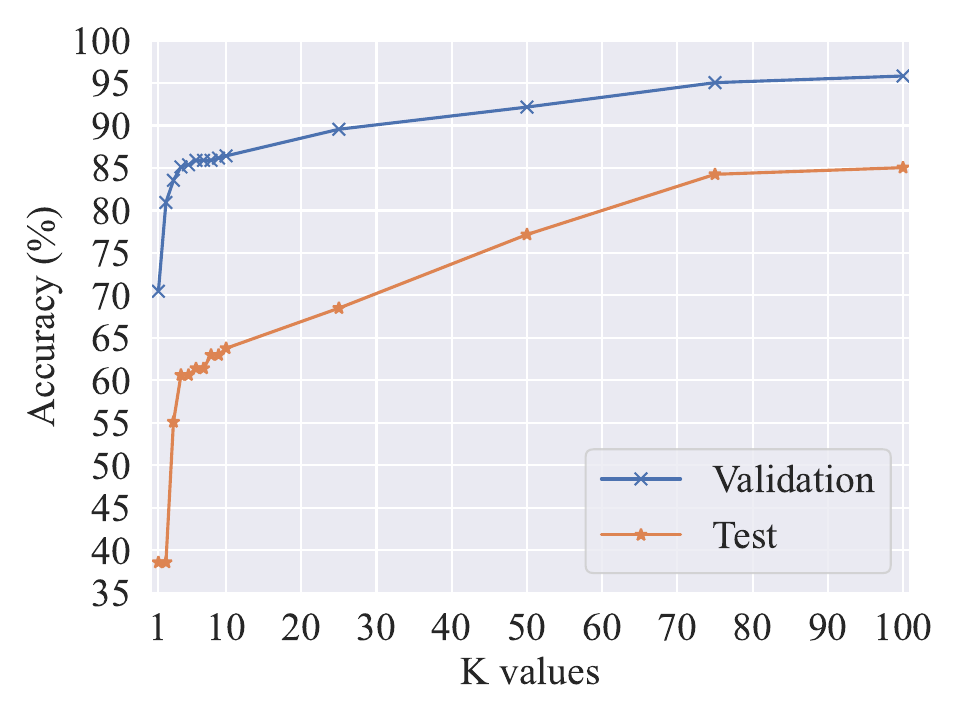}}
    \caption{The top-K accuracy of the decision tree on the validation set and on the test flight.}
    \label{fig:topk_results}
\end{figure}

The data set was separated into $70\%$ for training, $30\%$ for validation.
Instead of composing a test set with shuffled examples from training data, which
would incur in data leakage due to the similarity of the channels, we generated
a separate test trajectory. On the validation and test trajectories the model
yielded the performance displayed in Fig.~\ref{fig:topk_results} and
TABLE~\ref{tab:topk_results}, which is compatible to the ones presented in the
literature for \ac{NLOS} beam selection \cite{wen2023vision}. The decision tree
yielded the best results using a maximum depth of $15$.

\begin{table}[htb]
    \centering
    \small
    \caption{Top-K accuracy obtained from the decision tree model}
    \begin{tabular}{c|c|c}
        \textbf{K values} & \textbf{Validation acc. (\%)} &
        \textbf{Test acc. (\%)}
        \\
        \hline
        $1$               & $70.49$                       & $38.58$ \\
        $2$               & $80.93$                       & $51.18$ \\
        $3$               & $83.55$                       & $55.11$ \\
        $4$               & $85.11$                       & $60.62$ \\
        $5$               & $85.37$                       & $60.62$ \\
        $10$              & $86.42$                       & $63.77$ \\
        $25$              & $89.55$                       & $68.50$ \\
        $50$              & $92.16$                       & $77.16$ \\
        $75$              & $95.03$                       & $84.25$ \\
        $100$             & $95.82$                       & $85.03$
    \end{tabular}
    \label{tab:topk_results}
\end{table}

The top-K accuracy is the percentage of times that the model inference was found
in the K best beams for each specific step. So, the model was able to being at
least between the $100$ best choices in $85\%$ of the time, during the \ac{SAR}
mission. In the case of a sweeping being executed only on this reduced search
space, it would represent a reduction in the order of approximately $60.9\%$ of
the original full beam sweeping search space of $256$ possible beam indexes.

\subsection{Search and Rescue Mission}
\label{subsec:sar_mission}
Next, this same trained decision tree model was used to choose between beam
pairs for the process of beam selection in a \ac{SAR} mission, as described in
Subsection~\ref{subsec:dec_tree}. In this mission, the object detection model is
having its ability of detecting and identifying the rescue target using images
obtained by the \ac{UAV} bound to the service provided by the \ac{BS}.

The \ac{SAR} mission adopted two criteria: the first being the total mission
time in seconds, and the second being the quantity of successfully rescued
people. So, the decision tree and the two baseline approaches when used for this
\ac{SAR} mission yielded the performance described in
TABLE~\ref{tab:sar_mission_results}.

\begin{table}[htb]
    \centering
    \small
    \caption{Performance of each approach on the rescue mission}
    \begin{tabular}{c|c|c}
        \textbf{Approach}      & \textbf{\begin{tabular}[c]{@{}c@{}}Total
                                                 mission time \\
                                                 (seconds)\end{tabular}} &
        \textbf{\begin{tabular}[c]{@{}c@{}}Rescued \\
                        targets\end{tabular}}                              \\
        \hline
        \textbf{Random}        & $424.94$                                 & $3$
        \\
        \textbf{Decision tree} & $98.50$                                  & $5$
        \\
        \textbf{Oracle}        & $75.53$                                  & $5$
    \end{tabular}
    \label{tab:sar_mission_results}
\end{table}

As expected, the decision tree had the intermediate performance, between the
random and the oracle approaches. However, given that the oracle is the ideal
performance and therefore used only for reference, and the random one allows for
not only a large delay in the mission time but also misses two people, the
decision tree was capable to validate the methodology.

\subsection{\ac{CAVIAR} Benchmarking}
\label{subsec:benchmarking}

\begin{table}[htb]
    \centering
    \small
    \caption{Simulation hardware specifications}
    \begin{tabular}{c|c}
        \textbf{CPU}     & Intel® Core™ i7-8700 CPU @ 3.20GHz x 12 \\
        \hline
        \textbf{RAM}     & 32.0 GiB                                \\
        \hline
        \textbf{GPU}     & NVIDIA GeForce RTX 2060 Rev.A           \\
        \hline
        \textbf{Storage} & SSD WD blue SATA 3.0                    \\
    \end{tabular}
    \label{tab:machine_specs}
\end{table}

\begin{figure*}[htb!]
    \center
    \subfloat[Average CPU consumption.  \label{fig:cpu_comsump_1uav}]{%
        \includegraphics[width=0.81\textwidth]{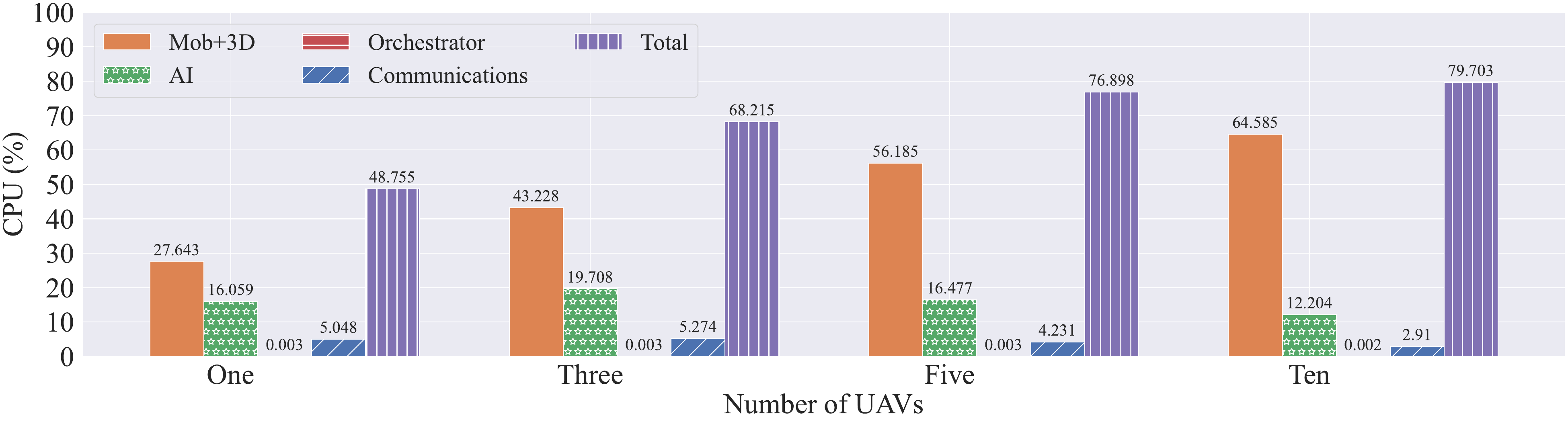}}\\
    \subfloat[Average RAM consumption. \label{fig:gpu_cpu_comsump_2uav}]{%
        \includegraphics[width=0.81\textwidth]{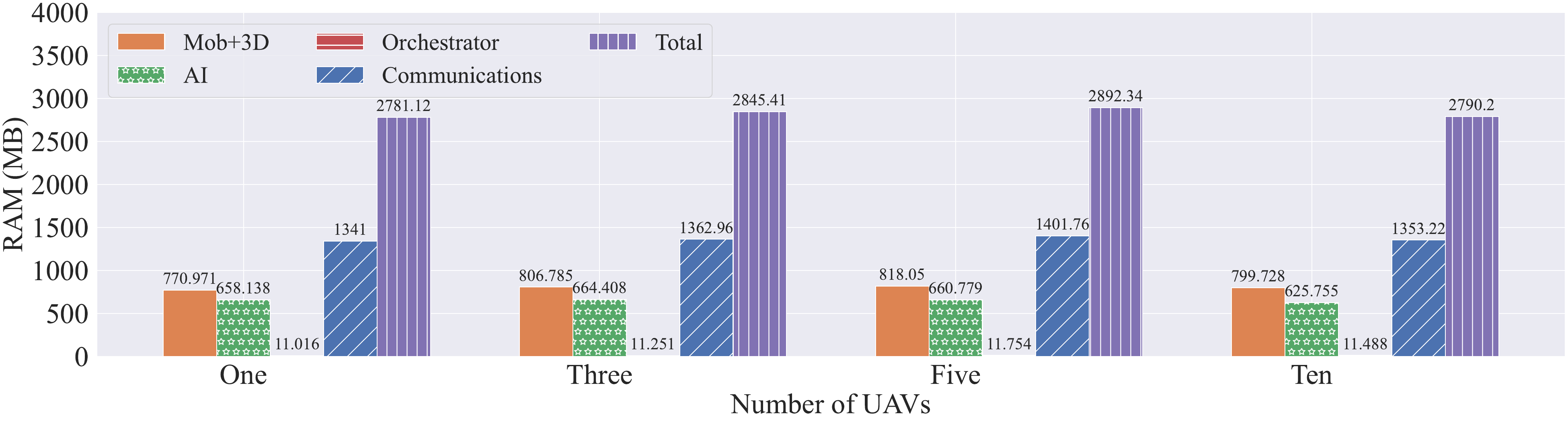}}\\
    \subfloat[Average GPU processing consumption.
        \label{fig:gpu_cpu_comsump_5uav}]{%
        \includegraphics[ width=0.81\textwidth]{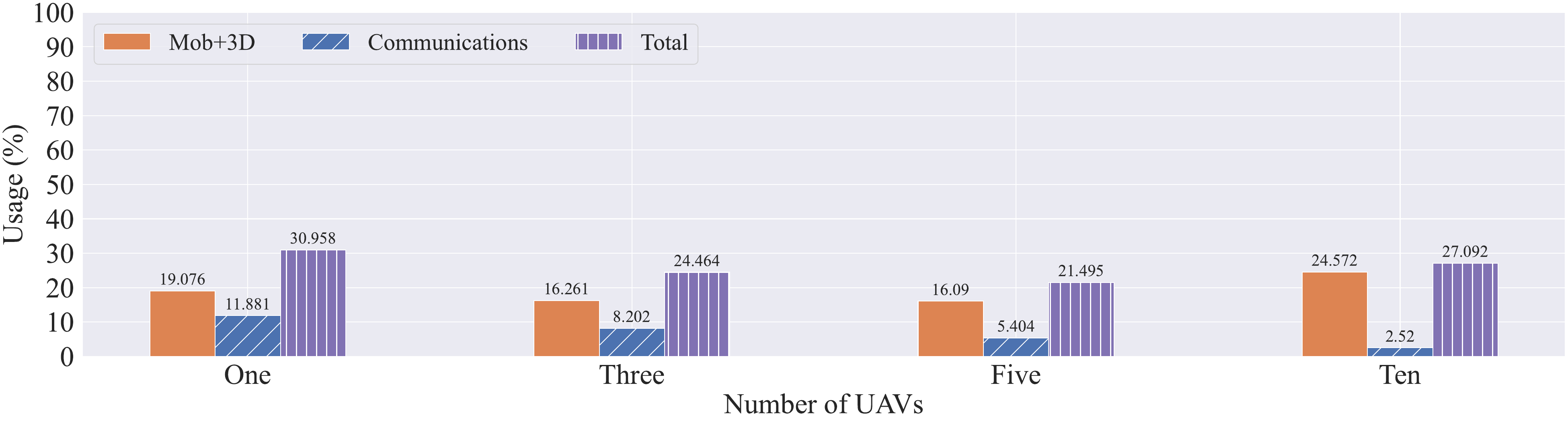}}\\
    \subfloat[Average GPU memory consumption. \label{fig:gpu_ram_comsump_1uav}
    ]{%
        \includegraphics[width=0.81\textwidth]{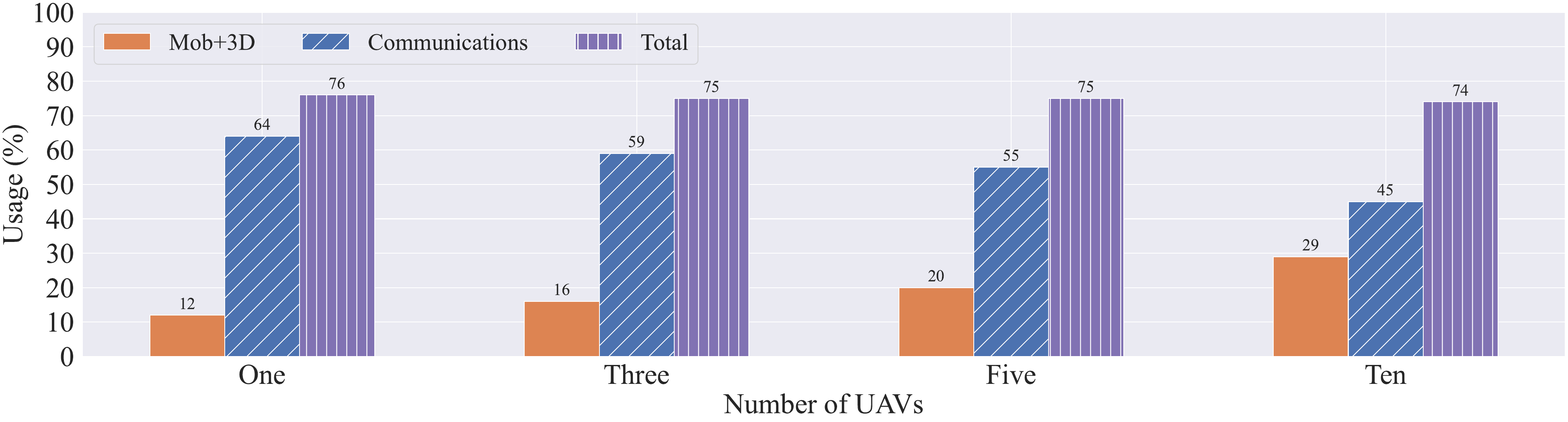}}\\
    \caption{Computational resource usage in terms of total CPU (the \%
        considers all $12$ CPUs), RAM, GPU processing, and GPU memory
        comsumptions for scenarios with $1$, $3$, $5$, and $10$ UAVs receivers.
        Each UAV is instantiated as a smart vehicle in AirSim, the 3D module,
        and as a receiver in Sionna, the Communications module.}
    \label{fig:cpu_ram_benchmarking_results}
\end{figure*}

\begin{figure}[htb]
    \centerline{\includegraphics[scale=.45]{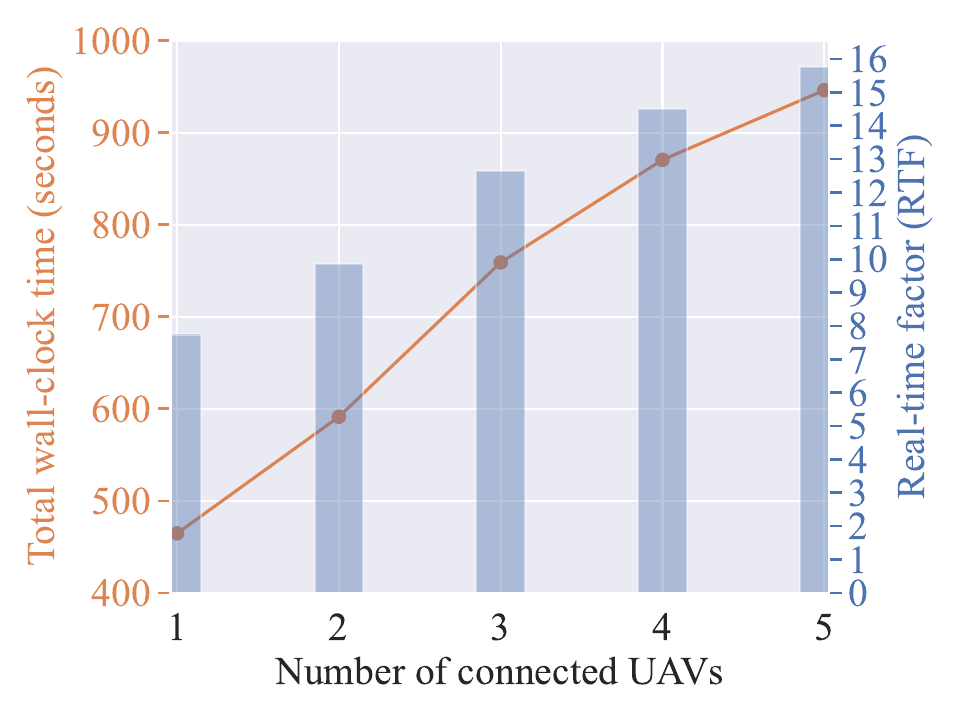}}
    \caption{Total wall-clock time for executing $60$ seconds of simulation without executing any rescue, and considering a sampling interval of $500$ ms.}
    \label{fig:simulation_times}
\end{figure}

Finally, to demonstrate the CAVIAR computational resource usage and its
feasibility, a benchmarking experiment was performed. For this, the first $60$
seconds (virtual time) of the flight mission from
Subsection~\ref{subsec:sar_mission} was executed four times, each one with the
instantation of $1$, $3$, $5$, and $10$ \acp{UAV} respectively. However, this
time no rescue was executed, but all instantiated \acp{UAV} were actively
capturing and sending images to the object detection \ac{AI} for rescue
identification. This is to avoid longer testing times, as executing the rescue
leads to moments were the drone is idle, which in turn leads to less resources
consumption during the $60$ seconds analyzed.

During the experiment, the consumption of the following resources were stored
for each module: 1) CPU, 2) RAM, 3) GPU processing, and 4) GPU memory.

Also, the resource usage was measured, together with the total time it took to
complete the $60$ second mission, Fig.~\ref{fig:simulation_times} shows the
total wall-clock time for each simulation when increasing the number of UAVs
present. The measurements of CPU, RAM, and GPU are displayed in
Fig.~\ref{fig:cpu_ram_benchmarking_results}. The specifications of the machine
used to perform this experiment are shown in TABLE~\ref{tab:machine_specs}.
Given that $T_s=500$~ms, as adopted in Fig.~\ref{fig:simulation_times}, is too
long for many applications in \ac{5G/6G}, significant efforts are required to
reduce the computational cost of critical simulations.

From Fig.~\ref{fig:cpu_ram_benchmarking_results} the results suggest that the
approach scales well in terms of RAM usage. Note how the CPU processing for the
3D module appears as the main increase in resource usage, while the RAM and GPU
processing allocated remains constant and relatively low, with the exception of
the GPU memory that, although constant, shows a high usage of around $75$\%.

The CPU consumption tends to increase when simulating more vehicles in the 3D
module. However, the main limitation currently is the number of receivers
simulated by the communications module. The limit is up to $10$ AirSim vehicles,
associated with a \ac{Rx} in Sionna. Adding another receiver results in needing more
GPU memory than available in the setup from TABLE~\ref{tab:machine_specs} to
execute the necessary ray tracing simulations. This can be alleviated by
limiting the number of intelligent vehicles simulated (i.e. vehicles
instantiated by AirSim). For example, using just $1$ AirSim controlled vehicle,
one can add another $13$ users not controlled by AirSim, having $14$ in total.
However, when a given use case need a large number of users, a less resource
hungry alternative for channel calculation needs to be used, such as stochastic
channel models. Another possibility is to distribute the simulation through more
than one machine, which is enabled by the current orchestrator implementation,
but further exploring these alternatives go out of the scope of the current
work.

\section{Conclusions}
\label{sec:conclusions}
The emergence of digital twins systems as a significant technology for future
communication networks calls for a robust and flexible simulation approach in
order to effectively implement virtual twins. For this, a potential option is to
use co-simulation methods. However, many multidisciplinary simulation
methodologies are bound to specific tools and use cases, lacking the flexibility
needed for \acp{DT}, that are usually tailor-made for each problem, therefore
constraining the development of \acp{VTwin} into a case-by-case approach and
hindering further improvements. This work builds upon the \ac{CAVIAR}
co-simulation methodology, implementing a modular approach that allows for
future upgrades and improved maintenance. This paper also showed benchmarking
experiments and data, proving that the approach can run in a single personal
computer with reasonable computing power. Results suggest that, for the present
\ac{SAR} use case, the approach scales well with the addition of more users,
with the exception being when smart vehicles are instantiated in the 3D
autonomous vehicles simulator, which quickly increases the CPU processing and
GPU memory usage.

\ac{CAVIAR} now supports the generation of experiments that closely integrate,
in an all-in-loop manner, \ac{AI}, communications, and 3D simulations, while
being flexible enough for other use cases, such as acting as the \ac{VTwin} of a
digital twin system. Due to the diversity of use cases in \ac{5G/6G} \acp{DT},
the list of future features is long. An important future work consists on
providing results with the \ac{PTwin} in the loop, by integrating the current
methodology with a real life network. Another ongoing work is to show results of
end-to-end simulations, by running Sionna alongside ns-3/5G-LENA. Minor
improvements include incorporating energy consumption tracking for the \ac{UAV},
enhancing 3D models in order to better create multi-modal information (RGB,
LiDAR, etc.), and improve support for addition of different 3D scenarios,
automatically adjusting their resolutions for better \ac{RT} execution time.
Furthermore, we also envision executing distributed simulations, and
experimenting with \ac{DT} composition by interacting \ac{DT} networks of
different network domains.

\bibliographystyle{IEEEtran}
\bibliography{references.bib}
\begin{acronym}
    \acro{CACC}{{Cooperative Adaptive Cruise Control}}
    \acro{GAN}{{Generative Adversarial Network}}
    \acro{OSM}{{OpenStreetMap}}
    \acro{JSON}{{JavaScript Object Notation}}
    \acro{ISAC}{{Integrated Sensing and Communications}}
    \acro{RIS}{{Reconfigurable Intelligent Surface}}
    \acro{PoC}{{Proof-of-concept}}
    \acro{PSNR}{{Peak signal-to-noise}}
    \acro{NLOS}{{Non-Line-of-Sight}}
    \acro{LOS}{{Line-of-Sight}}
    \acro{DT}{{Digital Twin}}
    \acro{PTwin}{{Physical Twin}}
    \acro{VTwin}{{Virtual Twin}}
    \acro{CGI}{{Computer-Generated Imagery}}
    \acro{CARLA}{{Car Learning to Act}}
    \acro{NATS}{{Neural Autonomic Transport System}}
    \acro{CAD}{{Cooperative Automated Driving}}
    \acro{5G/6G}{{fifth and sixth generation of mobile wireless technologies}}
    \acro{AI/ML}{{Artificial Intelligence/Machine Learning}}
    \acro{ML}{{Machine Learning}}
    \acro{DNN}{{Deep Neural Network}}
    \acro{UAV}{{Unmanned Aerial Vehicle}}
    \acro{BS}{{Base Station}}
    \acro{MIMO}{{Multiple Input Multiple Output}}
    \acro{SAR}{{Search and Rescue}}
    \acro{AI}{{Artificial Intelligence}}
    \acro{UE}{{User Equipment}}
    \acro{VM}{{Virtual Machine}}
    \acro{RIC}{{RAN Intelligent Controller}}
    \acro{CAVIAR}{{Communication Networks, Artificial Intelligence and Computer Vision with 3D Computer-generated Imagery}}
    \acro{VR}{{virtual reality}}
    \acro{DFT}{{Discrete Fourier Transform}}
    \acro{KPI}{{Key Performance Indicator}}
    \acro{MPC}{{Multipath Component}}
    \acro{RT}{{ray tracing}}
    \acro{Rx}{{receiver}}
    \acro{SUMO}{{Simulation of Urban MObility}}
    \acro{Tx}{{transmitter}}
    \acro{WI}{{Remcom's Wireless InSite®}}
    \acro{YOLO}{{You Only Look Once}}
    \acro{ns-3}{{Network Simulator 3}}
    \acro{0MQ}{{ZeroMQ}}
\end{acronym}
\end{document}